\documentclass[apjl,twocolumn]{emulateapj}

\usepackage{natbib}
\usepackage{amsmath}
\usepackage{hyperref}
\usepackage{subfigure}

\newcommand{\mcmc}[1]{#1}
\newcommand{\revision}[1]{#1}
\newcommand{\revagain}[1]{#1}

\begin{document}

\title{Thermal Diagnostics with the Atmospheric Imaging Assembly \\ onboard the Solar Dynamics Observatory: A Validated Method for\\ Differential Emission Measure Inversions}
\shorttitle{Thermal Diagnostics with SDO/AIA: A Validated Method for DEMs}

\author{Mark C. M. Cheung\altaffilmark{1},  P. Boerner\altaffilmark{1}, C. J. Schrijver\altaffilmark{1}, P. Testa\altaffilmark{2}, F. Chen\altaffilmark{3}, H. Peter\altaffilmark{3}, A. Malanushenko\altaffilmark{1,4}}

\affil{1. Lockheed Martin Solar and Astrophysics Laboratory, \\3251 Hanover Street Bldg. 252, Palo Alto, CA 94304, USA}
\affil{2. Harvard-Smithsonian Center for Astrophysics, \\60 Garden Street, Cambridge, MA 02138, USA}
\affil{3. Max Planck Institute for Solar System Research, \\Justus-von-Liebig-Weg 3, 37077 G\"ottingen, Germany}
\affil{4. High Altitude Observatory, National Center for Atmospheric Research,\\ 3080 Center Green Drive, Boulder, CO 80301, USA}

\email{cheung@lmsal.com}

\begin{abstract}
We present a new method for performing differential emission measure (DEM) inversions on narrow-band EUV images from the Atmospheric Imaging Assembly (AIA) onboard the Solar Dynamics Observatory (SDO).~\mcmc{The method yields positive definite DEM solutions by solving a linear program.} This method has been validated against a diverse set of thermal models of varying complexity and realism. These include (1) idealized gaussian DEM distributions, (2) 3D models of NOAA Active Region 11158 comprising quasi-steady loop atmospheres in a non-linear force-free field, and (3) thermodynamic models from a fully-compressible, 3D MHD simulation of AR corona formation following magnetic flux emergence. We then present results from the application of the method to AIA observations of Active Region 11158, comparing the region's thermal structure on two successive solar rotations. Additionally, we show how the DEM inversion method can be adapted to simultaneously invert AIA and XRT data, and how supplementing AIA data with the latter improves the inversion result. The speed of the method allows for routine production of DEM maps, thus facilitating science studies that require tracking of the thermal structure of the solar corona in time and space.
\end{abstract}


\maketitle

\section{Introduction}



Since the launch of NASA's Solar Dynamics Observatory~\citep[SDO;][]{Pesnell:SDO} in 2010, the
 Atmospheric Imaging Assembly~\citep[AIA; ][]{Lemen:AIA,Boerner:AIA} instrument onboard SDO has been delivering EUV imaging observations of the solar corona with an unprecedented combination of temperature coverage, spatial resolution, cadence and consistency in data quality.


AIA's simultaneous use of multiple spectral bands spanning the range of coronal temperatures also promises the ability to diagnose the thermal evolution of the observed systems.  However, while the instrument is, by design, well-suited for constraining the temperature of optically-thin plasma along its line of sight, the routine interpretation of AIA data in terms of the temperature and density of the emitting plasma remains a difficult problem to solve. 
Because the AIA EUV channels have temperature response functions that are generally multithermal~\citep[i.e. they have contributions from plasma over a range of temperatures, see][]{MartinezSykora:ForwardModelingAIA,Boerner:AIA}, the thermal distribution of the observed plasma cannot be directly inferred from the EUV images. To learn about coronal temperatures, one must separate the multithermal nature of coronal plasma from the multithermal response of the EUV channels.


In the many studies that have used AIA data to attempt to infer thermodynamic information about the corona, it has often been necessary to use forward fitting or qualitative comparisons with the observed multi-spectral characteristics of the plasma. We have developed a technique to utilize AIA data to their full potential for probing the thermal structure of the corona by producing maps of the differential emission measure DEM$(x,y,\log T,t)$ at any scale, up to and including the full cadence and spatial resolution of the AIA instrument. The method we present in this paper has been validated against a diverse set of thermal coronal models of varying sophistication and realism.

The remainder of the article is structured as follows. In section~\ref{sec:statement}, we formulate the mathematical problem underlying DEM analysis and discuss the requirements for any inversion method.  In section~\ref{sec:sparse_method}, we present our novel method for solving the inversion problem. In section~\ref{sec:validation}, we test the method against three different classes of DEM models. These include log-normal DEMs, distributions from quasi-steady loop models, and distributions from a fully-compressible, time-dependent MHD model of AR corona formation. In section~\ref{sec:aia_data}, we apply our DEM method to study the thermal distribution of an active region observed by AIA. In section~\ref{sec:aia_xrt}, we investigate the benefits of performing joint DEM inversions using narrowband AIA EUV and broadband X-ray imaging data from the X-Ray Telescope~\citep[XRT,][]{Golub:XRT} onboard Hinode~\citep[][]{Kosugi:Hinode}. A discussion of the implications and scientific possibilities resulting from this work is given in section~\ref{sec:discussion}.

 

\section{Statement of the problem}
\label{sec:statement}
\subsection{DEMs from optically thin emission}
\label{sec:dems}
Narrowband EUV (or broadband X-ray) observations can be related to the physical properties of optically thin coronal plasma as an integral over temperature space:
\begin{equation}
y_i = \int_0^\infty K_i (T)~{\rm DEM}(T) dT,\label{eqn:yintegral}
\end{equation}
\noindent where $y_i$ is the exposure time-normalized pixel value in the $i$-th AIA channel (in units of DN s$^{-1}$ pixel$^{-1}$) and $K_i(T)$ is the temperature response function (in units of DN cm$^{5}$ s$^{-1}$ pixel$^{-1}$, see Fig.~\ref{fig:response}).The differential emission measure (in units of cm$^{-5}$ K$^{-1}$) is defined by the relation DEM$(T)dT = \int_0^\infty n_e^2(T) dz$, where $n_e(T)$ is the electron number density of plasma at a certain temperature $T$.  The integral of DEM$(T)$ over a finite temperature range is simply called the emission measure (EM). The aim here is to take EUV imaging observations from AIA and invert for the emission measure distribution in the solar corona.~\revision{Although the functions $K_i$ are most sensitive to changes in temperature, they also depend on either the electron density or pressure. Furthermore, they depend on the choice of atomic abundances. The choices we made for this work are given in section~\ref{sec:matrix}.} 

The DEM inversion problem formulated above is an example of Fredholm's integral equation of the first kind~\citep[see, e.g.][]{CourantHilbert:MathematicalPhysics,Phillips:IntegralEquations1stKind}. In this mathematical framework, the measurement vector $\vec{y}$ is an integral transform of DEM$(T)$. \citet{CraigBrown:LimitationsOfSpectra}\footnote{\citet{CraigBrown:LimitationsOfSpectra} applied their analysis to DEM inversions from X-ray line spectra, but the same conclusions apply to DEMs from broadband observations.}  identified a number of key concerns for DEM analysis that result from well-known mathematical properties of this type of integral equation:
\begin{itemize}
\item{Given $\vec{y}$, there may be no solution for DEM$(T)$;}
\item{Even if a solution exists, it may not be unique;}
\item{Even if a solution exists, it may be unstable in the sense that small changes in $\vec{y}$ result in large changes in DEM$(T)$. This implies measurement errors in $\vec{y}$ may be amplified in the DEM solution;}
\item{Even if a solution exists, it may not be positive (semi)definite.}
\end{itemize}
\noindent These concerns highlight the potential pitfalls of DEM inversion analysis. For practical purposes, the first may not be the most pressing concern. If no solution exists for a measurement vector $\vec{y}$, it probably implies the assumed physical model is inappropriate. For instance, the emitting plasma may not be optically thin, or it may have an atomic abundance that is different to that used for computing the response functions $K_i(T)$. Or the plasma may have evolved in between exposures in different channels such that no single DEM$(\log T)$ satisfies Eq. (\ref{eqn:yintegral}) for all channels.

The remaining concerns motivate the following requirements for any DEM inversion scheme:
\begin{enumerate}
\item{The scheme needs a deterministic way to pick a solution out of a family of possible solutions that satisfy Eq. (\ref{eqn:yintegral}), and the chosen solution should be representative of the emitting plasma;}
\item{The solution returned by the inversion scheme should be stable. That is, for a noisy measurement $\vec{y}+\vec{e}$, where $\vec{y}$ represents the noiseless measurement and $\vec{e}$ represents random errors, the inversion scheme needs to return similar DEM solutions over different realizations of $\vec{e}$;}
\item{The solution returned should be positive semidefinite, i.e. DEM$(T)\ge 0$.}
\end{enumerate}
\noindent An additional desired property of inversion schemes is computational speed. This is especially true for DEM analysis of AIA data. In normal operational mode, AIA delivers a complete set of seven EUV images every 12 seconds. This corresponds to $\sim 10^6$ observation vectors ($\vec{y}$'s) per second. Even when subsampling and/or averaging (either spatially or temporally) is used to reduce throughput, it would still be desirable for a DEM inversion code to be able to return at least $10^4-10^5$ solutions per second. 


\subsection{Matrix formulation of the problem}
\label{sec:matrix}
Different DEM inversion methods reported in the literature have different ways to satisfy (or not) the requirements listed in the previous section. It is instructive to formulate the DEM inversion problem in matrix form to facilitate further discussion.  In practice, the integrals in Eq.~(\ref{eqn:yintegral}) are always approximated as a sum over discrete points. Eq.~(\ref{eqn:yintegral}) can then be expressed as a set of linear integral equations~\citep[e.g. see][]{CraigBrown:LimitationsOfSpectra}. After a quadrature scheme is chosen (i.e.~in terms of basis functions, see Appendix~\ref{sec:quadrature}), the set of equations can be written in the form:
\begin{equation}
\vec{y} = \mathbf{D}\vec{x}.\label{eqn:system}
\end{equation}
\noindent The meanings of the matrix elements $D_{ij}$ and components of the solution vector $x_j$ depend on the quadrature scheme chosen to approximate the integral.

One choice of a quadrature scheme is the following. Let the logarithmic temperature range be divided into $n$ neighboring bins. Then Eq.~(\ref{eqn:yintegral}) becomes
\begin{equation}
y_i = \sum_{j=1}^{n} \int_{\log T_j}^{\log T_j+\Delta \log T_j} K_i(\log T) {\rm DEM}(\log T) d\log T,
\end{equation}
\noindent where the $j$-th temperature bin has range $\log T\in [\log T_j, \log T_j + \Delta \log T_j)$. Assume that $K_i(\log T) = K_{ij}$ is piecewise constant in each $j$-th temperature bin, so 
\begin{eqnarray}
y_i &=& \sum_{j=1}^{n} K_{ij}{\rm EM}_j,{\rm where}\label{eqn:linear_system}\\
{\rm EM}_{j} & = & \int_{\log T_j}^{\log T_j+\Delta \log T_j} {\rm DEM}(\log T)  d\log T.\label{eqn:linear_system_line2}
\end{eqnarray}
\noindent Thus Eq.~(\ref{eqn:yintegral}) has be transformed into the matrix Eq.~(\ref{eqn:linear_system}). In this quadrature scheme, $x_j = {\rm EM}_j$ so the components of the solution vector are simply values of the emission measure contained in discrete, non-overlapping temperature bins. $\mathbf{K}$ is an $m\times n$ matrix with components $K_{ij}$, $\vec{y}$ is an $m$-tuple corresponding to measurements by the AIA EUV channels. In practice, we chose to use a set of basis functions (including both Dirac-delta and Gaussians functions) to describe the DEM solution.~\revagain{As detailed in Appendix~\ref{sec:quadrature}, we use a set of Dirac-delta functions and three sets of Gaussians of different widths. In total $84$ basis functions are used for a $\log T$ grid with $21$ grid points spanning $\log T/K = 5.5$ to $\log T/K = 5.5$ at intervals of $\Delta \log T/K = 0.1$. In this case a slightly modified linear system is solved but the discussion below still applies.}

\begin{figure}
\centering \includegraphics[width=0.48\textwidth]{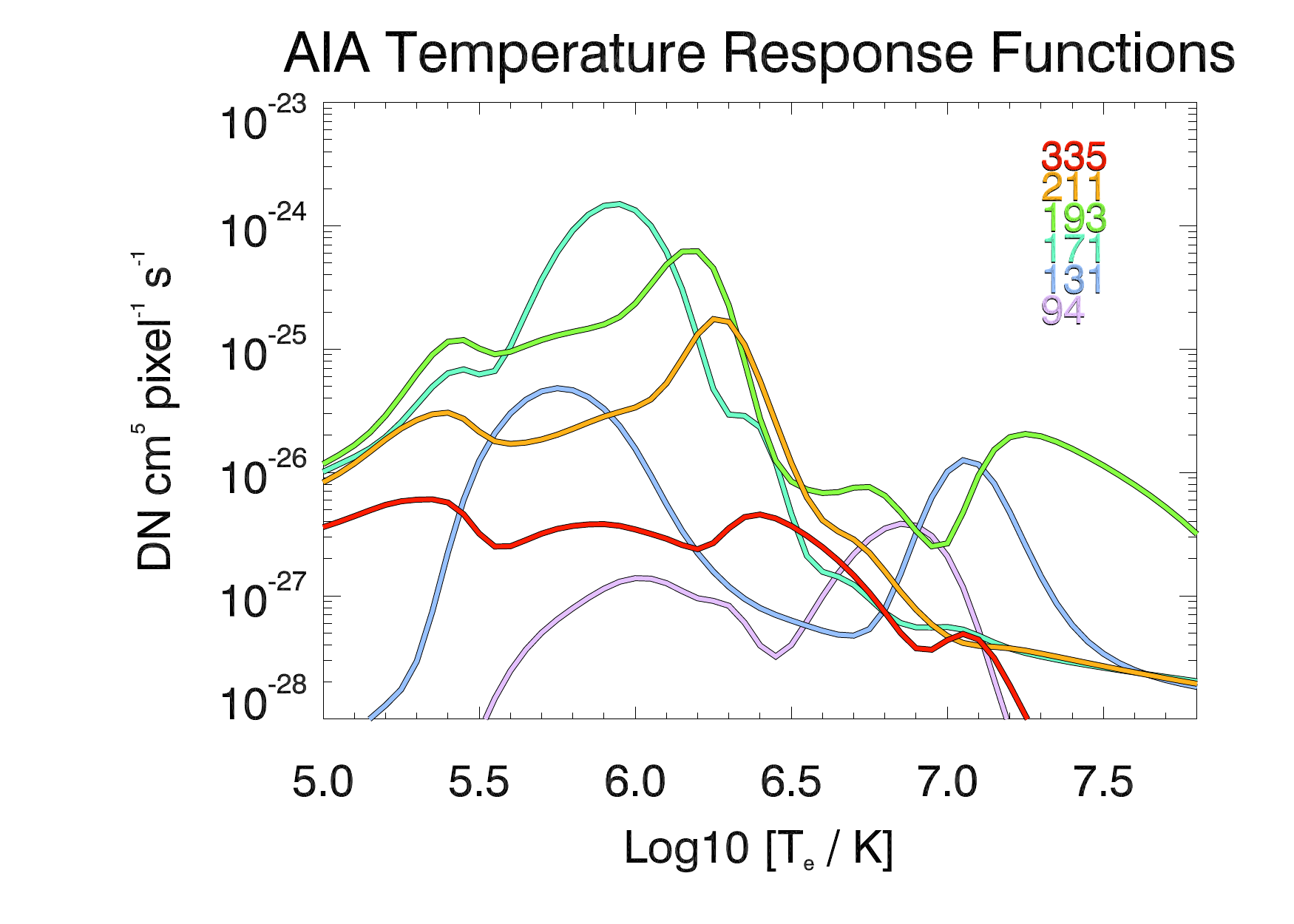}
\caption{Theoretical response functions $K(T)$ for the $94$, $131$, $171$, $193$, $211$ and $335$ \AA~EUV channels of SDO/AIA. The response functions were computed using CHIANTI 7.1.3~\citep{CHIANTI7.1}.}
\label{fig:response}
\end{figure}



Fig.~\ref{fig:response} shows temperature response functions for the AIA EUV channels as computed by the Solarsoft routine~\texttt{aia\_get\_response($/$temp,$/$evenorm,$/$chiantifix)} using the CHIANTI 7.1.3 package~\citep{CHIANTI7.1}. \revision{Coronal abundances specified in the~\texttt{sun\_coronal\_ext.abund} file~\citep[a compilation of abundances from][]{Feldman:1992,GrevesseSauval:StanfordSolarComposition,LandiFeldmanDere:2002} of the CHIANTI package were used and the pressure was set at $p/k = 10^{15}$ K cm$^{-3}$, where $k$ is the Boltzmann constant.} \revision{The \texttt{chiantifix} keyword applies an empirical correction to the temperature response function of the 94~\AA~channel~\citep[][]{Boerner:Crosscalibration} to account for missing transitions in the CHIANTI database.\footnote{\revagain{Older versions also applied an empirical correction to the 131~\AA~channel but this is no longer the case since version 6 of the AIA calibration.}}}

Although AIA has seven EUV channels, emission in the $304$ \AA~channel is often optically thick and is not well-modeled by CHIANTI under the optically thin assumption.  So for AIA DEM analysis the $94$, $131$, $171$, $193$, $211$ and $335$~\AA~channels are typically used (i.e. $m=6$). If the number of temperature bins were $n=m=6$, then one could in principle try to solve for $\vec{x}$ by multiplying both sides of Eq.~(\ref{eqn:linear_system}) by $\mathbf{K}^{-1}$. This is the solution method examined by~\citet{CraigBrown:LimitationsOfSpectra}. As already summarized in section~\ref{sec:dems}, they pointed out a number of problems with this approach. In particular, any noise in $\vec{y}$ will be amplified in $\vec{x}$ and there is no guarantee that the solution will be positive definite.

For $n > m$ (i.e. more than $6$ temperature bins), Eq.~(\ref{eqn:system}) is underdetermined. This is a well-known problem in emission measure inversions.  Section \ref{sec:chi2_method} summarizes commonly used techniques to deal with this problem. In section~\ref{sec:sparse_method}, we present a new method we developed based on the concept of sparsity.

\subsection{Methods based on $\chi^2$-minimization}
\label{sec:chi2_method}
Most DEM inversion schemes in the literature are based on a reduced-$\chi^2$ approach. That is, the DEM solution is chosen to be one such that 
\begin{equation}
\chi^2(\vec{x}) = \sum_{i=1}^{m} \left(\frac{y_i - \sum_j K_{ij}x_j}{\delta y_i}\right)^2\label{eqn:chi2}
\end{equation}
\noindent is minimized. Here $\delta y_i$ is the uncertainty for the $i$-th channel. This approach is ideal for overdetermined systems (i.e. $n < m$) where it is known that no single model will reproduce all $n$ measurements (linear regression through three or more non-colinear points is one example).  However, for underdetermined systems it is subject to the perils of overfitting. To mitigate this, additional constraints are often added to the definition of $\chi^2$ using the method of Lagrange multipliers. The objective function then becomes $\chi^2(\vec{x}) + F(\vec{x})$. The regularization term $F(\vec{x})$ is used to penalize certain solutions with properties that are undesirable. 

A common choice for regularization is the smoothness constraint, which is imposed to reject solutions that exhibit oscillatory behavior~\citep[e.g.][]{Phillips:IntegralEquations1stKind,CraigBrown:InverseProblems,MonsignoriLandini:DEMs,HubenyJudge:1995,Judge:Inversions}. Smoothness in the optimal solution is sought by imposing a penalty term $F(\vec{x}) = \lambda \sum (x_{i-1}-2x_i+x_{i+1})^2$, where $\lambda$ is the Lagrange multiplier and the summand is a finite difference formula for the second derivative. This is sometimes called `second order' regularization.

Another choice is the so-called `zeroth order' regularization. In this scheme the penalty term is $\lambda||\vec{x}||^2_2 $, where $||\vec{x}||_2 = \sqrt{\sum_i x_i^2}$ is the L$2$-norm (i.e. the Euclidean norm) of $\vec{x}$. Zeroth-order regularization was recently used by~\citet{HannahKontar:AIADEMs} and~\citet{Plowman:FastDEMs} for DEM inversions of AIA data. As discussed by~\citet{Judge:Inversions} and~\citet{Plowman:FastDEMs}, zeroth-order regularization has direct correspondence with the singular value decomposition (SVD) approach to solving the underconstrained problem. Both return solutions that have the minimum total squared EM. It was pointed out by~\citet{Plowman:FastDEMs} that zeroth-order regularization is also a kind of smoothness constraint since, for the same emission measure, a sharply peaked solution will have higher total squared EM than a solution with a broader distribution.

Other regularization procedures are available~\citep[e.g. based on maximum entropy regularization, see][]{MonsignoriFossiLandi:DEMs,Judge:Inversions}. While regularization helps to mitigate the ill-conditioning of the DEM inversion problem, the problem remains that $\chi^2$-minimization schemes generally do not guarantee solutions that are positive definite. Instead, inversion schemes often resort to hybrid procedures that combine $\chi^2$-minimization with follow-up steps to mitigate negativity~\citep[e.g.][]{HannahKontar:AIADEMs,Plowman:FastDEMs}. 

One way to avoid solutions with negative values is to perform parametric inversions, i.e.~to enforce positive definite functional forms of solutions and then to perform $\chi^2$-minimization to find the optimal set of parameters. Commonly used functional forms of the DEM function include Gaussians~\citep{Aschwanden:AIADEMs,Guennou:AIADEMsI,Guennou:AIADEMsII}, power laws~\citep[][]{Jordan:SolarActiveRegions} or a combination of the two~\citep{Guennou:DepositionTimescale}. Discretized splines have also been used~\citep[][, see also the~\texttt{xrt\_iterative2} inversion code by~Weber, available in the Hinode/XRT package in Solarsoft]{MonsignoriFossiLandi:DEMs,Parenti:CoronalStreamers}.

A more comprehensive exploration of the space of possible DEM solutions can be undertaken using Markov-Chain Monte Carlo methods~\citep[MCMC;][]{KashyapDrake:MCMC}. These algorithms begin with a guess at the DEM and iteratively apply randomized adjustments in a Markov chain, producing a family of DEM solutions that can be thought of as a representation of the probability distribution function of the actual DEM. The MCMC method does not impose a pre-determined functional form for the DEM (though it applies some locally variable smoothness based on the shape and coverage of the temperature responses), and it is one of the few available methods that provides estimates of the uncertainties associated with the DEM. However, the MCMC method is computationally demanding, typically requiring seconds to minutes for each observation vector, therefore making it not particularly suitable for application to large AIA datasets with high spatial and temporal resolution.

\citet{Testa:DEMReliability} used 3D radiative MHD simulations of Bifrost~\citep{Gudiksen:Bifrost} to test the reliability of DEM diagnostics applying the MCMC method to AIA and Hinode/EIS synthetic data. When applying the MCMC method to AIA data they find that though the general features and spatial distribution patterns of DEMs can be reconstructed, there are some limitations: (a) the temperature at which the DEMs peaks is systematically slightly underestimated; (b) while isothermal DEMs are reasonably well reconstructed, inversion solutions for synthetic data generated by broad DEMs, especially those with significant density structuring along the l.o.s., were less accurate.



\section{A new method based on sparsity}
\label{sec:sparse_method}
We propose a new inversion method based on the concept of sparsity, which has received a great deal of attention in recent years by the compressed sensing community. Compressed sensing is concerned with the recovery of signals where the number of measurements is less than (sometimes much less than) the number of components in the reconstructed signals.

In an underdetermined linear system such as given by Eq.~(\ref{eqn:system}), the family of solutions satisfying the equation resides in an affine subspace of $\mathcal{R}^n$. The challenge is to select a solution within this subspace that most faithfully represents the underlying scenario. In a series of papers on solutions to underdetermined linear systems, \citet[e.g.][]{CandesTao2006:NearOptimal,CandesTao2007:DantzigSelector} showed that, when compared to a least-squares/minimum energy approach, the assumption of sparsity often results in a solution that is a better approximation to the real signal. This realization has led to immense advances in many fields where the reconstruction of a linear signal is desired from undersampled data~\citep[e.g. time series, images, and tomographic magnetic resonance imaging; see][]{Donoho:2006CompressedSensing,Lustig:SparseMRI}.

Mathematically, the most sparse solution is defined as the solution to the optimization problem:
\begin{equation}
{\rm minimize} ~|| \vec{x} ||_{0} {\rm~subject~to~} \mathbf{D}\vec{x} = \vec{y}.
\end{equation}
\noindent Here $|| \vec{x} ||_{0}$ is the L$0$ norm of $\vec{x}$, which is just the number of non-zero components of $\vec{x}$. There is no known efficient algorithm for solving this L$0$ norm minimization problem, so~\citet{CandesTao2006:NearOptimal} instead proposed that one should solve the corresponding L$1$ norm minimization problem, namely
\begin{equation}
{\rm  minimize} ~|| \vec{x} ||_{1} {\rm~subject~to~} \mathbf{D}\vec{x} = \vec{y},
\end{equation}
\noindent where $|| \vec{x} ||_{1} = \sum\limits_{j=1}^{n} || x_j ||$. This is the underpinning of our approach to tackling the EM inversion problem. 


In practice, systematic errors (e.g. in the instrument response matrix $K_{ij}$) and random errors in the measurement vector $\vec{y}$ means that the sought-after solution may not necessarily satisfy Eq.~(\ref{eqn:system}). Furthermore, for EMs we must impose that the solution be positive semidefinite (i.e. $x_j \ge 0$). So our method solves the following linear program:
\begin{eqnarray}
{\rm LP1: minimize} & \sum\limits_{j=1}^{n}  x_j &{\rm~subject~to}\\
 \mathbf{D}\vec{x} &\le& \vec{y} + \vec{\eta}, \label{eq:e1}\\
\mathbf{D}\vec{x} &\ge& {\rm max}(\vec{y} - \vec{\eta},0),\label{eq:e2}\\
\vec{x} &\ge& 0\label{eq:e3}.
\end{eqnarray}
\noindent The inequality constraint (\ref{eq:e3}) ensures the solutions are positive semidefinite. The inequality constraints (\ref{eq:e1}) and (\ref{eq:e2}) provide some tolerance for the solution to deviate from satisfying Eq.~(\ref{eqn:system}). 
\revision{In this paper, we set the tolerance level as $\eta_i = e_i$, where $e_i$ is an estimate of the uncertainty for a pixel count (DN/pixel) in the $i$-th EUV channel divided by the exposure time. $e_i$ can be computed using the function \texttt{aia\_bp\_estimate\_error}, which is available as part of the AIA package in Solarsoft. \texttt{aia\_bp\_estimate\_error} computes an estimated uncertainty for a given signal level in a given channel based on photon counting statistics and a number of instrumental effects: read noise, compression and quantization round-off, and error in dark subtraction.}

We are unaware of~\emph{physical} principles describing coronal plasmas that would motivate an objective function based on the L1 norm. However, this choice is appealing in a number of ways. First of all, this scheme minimizes the number of components (in terms of quadrature weights) needed to fit the observations, and in this sense it avoids the problem of overfitting. This behavior is consistent with the principle of parsimony (more commonly known as Ockham's Razor). Secondly, this scheme ensures positivity of the solution (if a solution is found).

Thirdly, the problem posed as LP1 lends itself to being solved by fast numerical techniques. The computational requirement of any DEM method is a practical concern since AIA delivers data at such a high rate (of order $10^5$ observation vectors $\vec{y}$ per second). The DEM inversion problem posed as LP1 is an example of~\emph{basis pursuit}. Basis pursuit is a technique commonly employed in the compressed sensing literature for reconstructing undersampled signals~\citep[][]{Chen:AtomicDecomposition}. Because we require $\vec{x}\ge 0$, the convex objective function $|| \vec{x}||_{1}$ reduces to the simple linear form $\sum_j x_j$. The linear program LP1 can then be solved efficiently using the simplex algorithm~\citep{Dantzig:Simplex}, which is designed to find optimal solutions to problems where the objective function is a linear form and the constraints are posed as linear inequalities. Our implementation of the DEM inversion code makes use of the~\texttt{simplex} function in the IDL data analysis package. The implementation of the simplex method in IDL is based on the method as detailed in section 10.8 of~\emph{Numerical Recipes}
by~\citet*{NumericalRecipes}. The computational speed of the inversion code is discussed in section~\ref{sec:speed}.

Regardless of the advantages listed above, a DEM inversion method would be worthless if it only (or mostly) returned solutions that are not representative of the emitting coronal plasma. In the next section, we present results from validation tests of the method.

\section{Validation Tests}
\label{sec:validation}
In this section, we test our inversion method against a diverse set of thermal models of varying complexity and realism. 

\subsection{Gaussian / log-normal DEM distributions}
\label{sec:gaussian_test}
\begin{figure*}
\includegraphics[width=\textwidth]{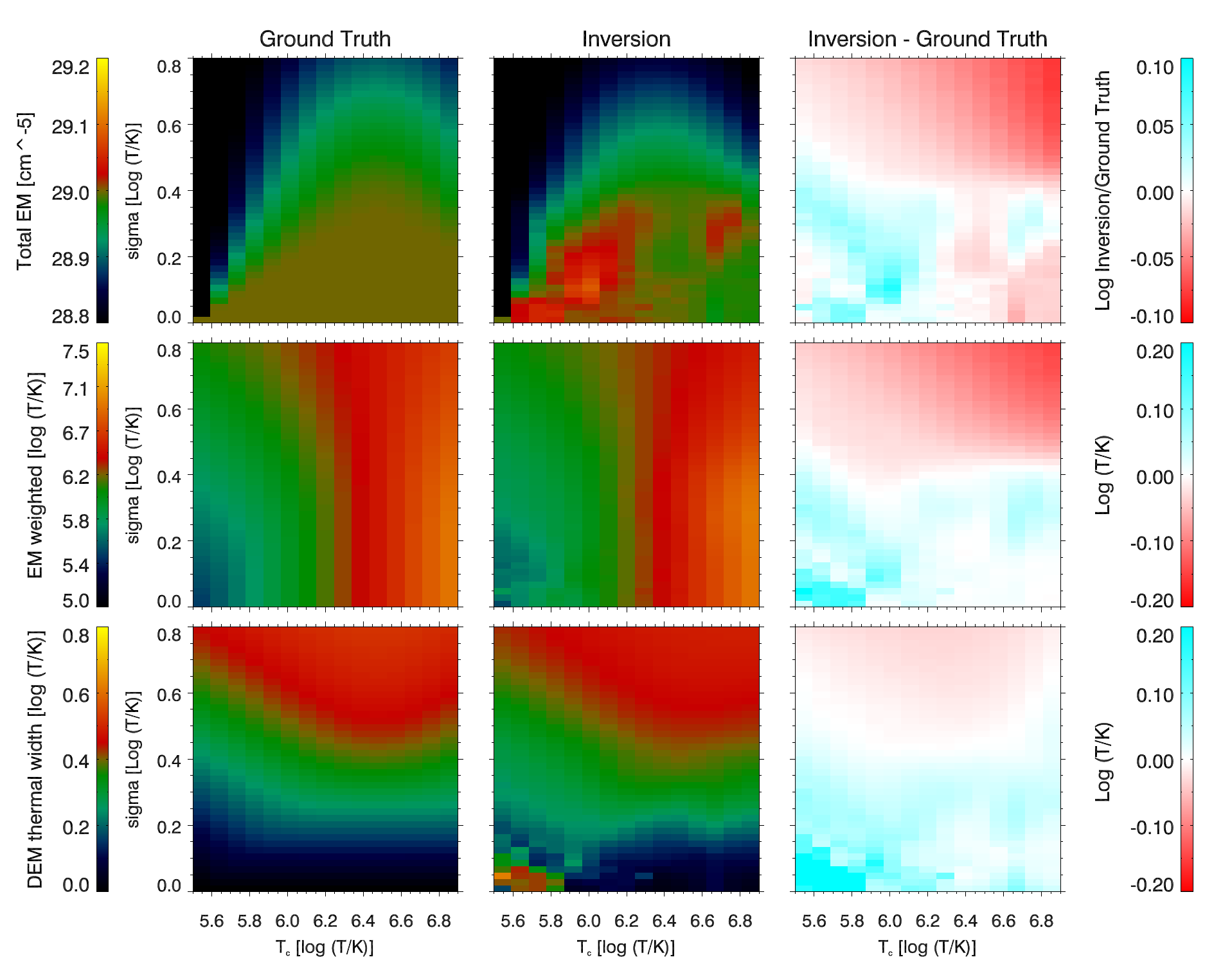}
\caption{Validation test on Gaussian DEM distributions. DEM inversions were carried out using synthetic count rates for six AIA EUV channels ($94$, $131$, $171$, $193$, $211$ and $335$). Noise was added to the synthetic count rates to generate an ensemble of noisy observation vectors for each model. The top, middle and bottom panels show ensemble averages of the three metrics used the quantify the fidelity of the DEM solutions (total EM, EM-weighted log T and thermal width) to the underlying DEM model.}\label{fig:gaussian_test}
\end{figure*}

\begin{figure*}
\includegraphics[width=\textwidth]{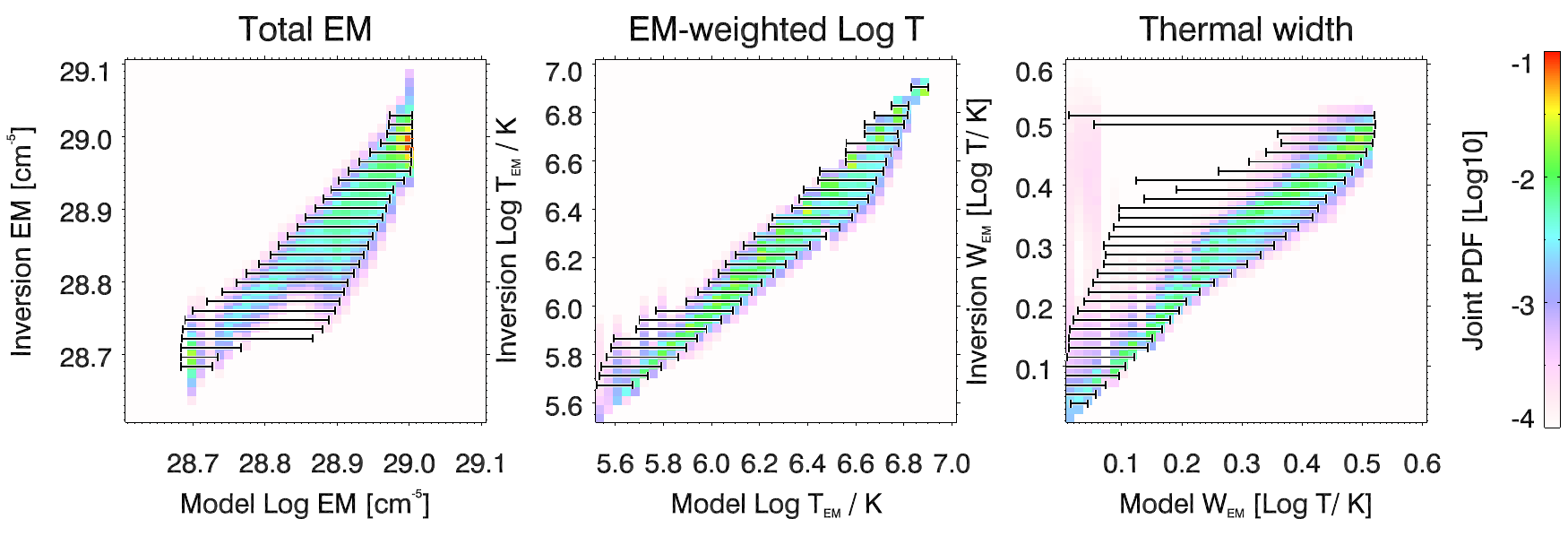}
\caption{Validation test on Gaussian DEM distributions. Probability density functions (PDFs) of EM, $T_{\rm EM}$ and $W_{\rm EM}$ as computed from the input DEM model (abscissa) and inversion solution (ordinate) are shown. The color-coding indicates the joint PDF $P(q^I , q^M)$, where $q^I$ and $q^M$ denote quantities from the inversion and model, respectively. The horizontal error bars indicate the $95\%$ confidence interval.}\label{fig:gaussian_jpdfs}
\end{figure*}

Log-normal distributions are commonly chosen to serve as test cases for inversion codes~\citep{HannahKontar:AIADEMs,Guennou:AIADEMsI,Guennou:AIADEMsII,Plowman:FastDEMs} and as functional forms for DEM inversions of AIA data~\citep[e.g.][]{Aschwanden:AIADEMs}. They correspond to Gaussian functions in $\log T$ space:
\begin{equation}
{\rm \xi}(T,T_c,\sigma) = \frac{{\rm EM}_0}{\sigma\sqrt{2\pi}}\exp{\left[-\frac{(\log T-\log T_c)^2}{2\sigma^2}\right]},
\end{equation}
\noindent where $T_c$ is the peak temperature and $\sigma$ is the Gaussian width. The normalization is chosen such that the total emission measure is EM$_0 = \int_0^\infty \xi  {\rm dlog}T $.\footnote{Strictly speaking, $\xi(T)$ is not the differential emission measure as defined in Eq.~(\ref{eqn:yintegral}). The two are related by the following relation DEM$(T)  = \ln{10} T^{-1} \xi(T)$. Nevertheless, we will follow the common practice in the literature and refer to both DEM$(T)$ and $\xi(T)$ as differential emission measure functions.} The validation test for the inversion method was performed over a parameter space of Gaussian distributions ranging from $\log T_c \in [5.5,7.0]$ and $\sigma \in [0.0,0.8]$. In the following, the total emission measure was set to EM$_0 = 10^{29}$ cm$^{-5}$.  

 For each $(T_c,\sigma)$ model, we computed $\xi(T)$ on a temperature grid spanning $\log T/K \in [5.5,7.5]$ at intervals of $\Delta \log{T} = 0.0025$. \revision{Note that for wide DEM distributions, the total EM contained within this temperature range can be significantly (up to $0.2$ dex) below $10^{29}$ cm$^{-5}$.}  We then folded $\xi$ with the AIA temperature response functions ${K}_i(T)$ to generate synthetic count rates (DN s$^{-1}$ pixel$^{-1}$) for the $94$, $131$, $171$, $193$, $211$ and $335$ channels. \revision{As discussed in section~\ref{sec:dems}, one of the potential pitfalls of DEM inversions is the amplification of observational noise in the inversion, which may render the solution unrepresentative of the underlying DEM.  To examine whether this is the case with the sparse inversion method}, we generated an ensemble of noisy observation vectors for each $(T_c,\sigma)$ model. The magnitude of the stochastic error $\vec{e}$ in the AIA channels was estimated using the Solarsoft function \texttt{aia\_bp\_estimate\_error} \revision{(as described in section~\ref{sec:sparse_method} for details)}. For each $(T_c,\sigma)$ model, we generated $5000$ instances of noisy observation vectors $y_j+\alpha_je_j$, where $\alpha_j$ is a random variable drawn from a normal distribution. We then performed the DEM inversion on every member of the ensemble. 

For the inversion, we used a temperature grid spanning $\log T/K \in [5.5,7.5]$, but at a much reduced grid spacing of $\Delta \log{T} = 0.1$. \revision{For a discussion of how the temperature grid was chosen, see Appendix~\ref{sec:tgrid}}. We have $n=21$ temperature bins for $m=6$ AIA channels. We define three metrics to quantify the fidelity of the inverted EM distribution vs. the input model:
\begin{eqnarray}
{\rm EM} &=& \sum_j^n  {\rm EM}_j , \\
\log T_{\rm EM}& =& {\rm EM}^{-1}\left[ \sum_j^n  {\rm EM}_j \log T_j \right],{\rm~and}\\
W_{\rm EM}^2  &=  &{\rm EM}^{-1}\left[ \sum_j^n  {\rm EM}_j (\log T_j - \log T_{\rm EM} )^2\right].
\end{eqnarray}
\noindent They correspond to the zeroth, first and second moments of the emission measure distribution. EM$_i$ denotes the emission measure contained in the $i$-th temperature bin and EM is the total emission measure. $\log T_{\rm EM}$ is EM-weighted $\log$ temperature and $W_{\rm EM}$ is the effective width of the distribution in $\log T$ space (indicating the extent of multithermality). We use the notation $\langle Q \rangle$ to denote the average of a quantity $Q$ over the ensemble (e.g. $\langle \log T_{\rm EM}\rangle$ is the average $\log T_{\rm EM}$ taken over the ensemble).

Figure~\ref{fig:gaussian_test} shows a comparison between the model and inverted DEMs. On average, the inversion scheme returns solutions that are representative of the underlying input models in terms of the metrics $\langle {\rm EM}\rangle$, $\langle \log T_{\rm EM} \rangle $ and $ \langle W_{\rm EM} \rangle $. Over the parameter space of $T_c$ and $\sigma$, the code is, in general, able to give total EMs with an error of less than $10-20\%$, $\log T_{\rm EM}$ within an error of $0.2 \log T/K$ and $W_{\rm EM}$ within an error of $0.2 \log T/K$.

Another way to examine the fidelity of inversion solutions with respect to the underlying model is to use joint probability density functions (joint PDFs, or 2D histograms). Fig.~\ref{fig:gaussian_jpdfs} shows joint PDFs for the three metrics EM, $T_{\rm EM}$ and $W_{\rm EM}$. For all three metrics, the majority of the solutions lie along the diagonal, indicating the inversion solutions are generally representative of the underlying DEM model. The horizontal error bars indicate the uncertainty of a given measurement from the inversion. Let $P(q^M | q^I) = P(q^M) [\int_0^\infty P(q, q^I) dq]^{-1}$ denote the conditional probability such that the underlying model DEM has metric value $q^M$ given $q^I$ from the inversion. Furthermore, let $C(q^M | q^I) = \int_0^{q^M} P(q | q^I) dq $ be the associated cumulative distribution function. The horizontal error bars in Fig.~\ref{fig:gaussian_jpdfs} span the range $C(q^M | q^I) = 0.025$ to $C(q^M | q^I) = 0.975$. In other words, given $T_{\rm EM}^I$, one can be $95\%$ confident that the underlying model DEM has $T_{\rm EM}^M$ within the range spanned by the error bars.

For the same Gaussian DEM model,~\citet{Guennou:AIADEMsII} examined the viability of using the six AIA EUV channels to constrain the DEMs of multithermal plasmas. For their multithermal DEM inversions, Gaussian solutions were assumed with EM, $T_c$ and $\sigma$ as parameters. Unregularized  $\chi^2$-minimization (i.e.~$F(\vec{x}) = 0$) was then performed to find the set of parameters that best reproduced the synthetic AIA counts. To ensure the minimum $\chi^2$ solution was found for each test case, they deployed a GPU-accelerated implementation of the inversion code and used a `brute force' approach to comprehensively scan the parameter space. Their systematic study raised concerns about the suitability of AIA data for DEM analysis. For Gaussian models with $\sigma=0.1 \log T/K$, they reported that $\chi^2$ minimization yielded, with high probability, optimal solutions with peak temperatures close to the underlying $T_c$ of the model. However, the fidelity of the inversions diminished with broader DEMs. For $\sigma = 0.7\log T/K$, there was a bias toward solutions with $T_c = 1$ MK for model peak temperatures ranging from $\log T_c / K = 5 - 7$. Also, the inverted value for $\sigma$ was only weakly correlated with the width of the underlying model for $\sigma \gtrsim 0.3\log T/K$.  

\begin{figure}
\includegraphics[width=0.5\textwidth]{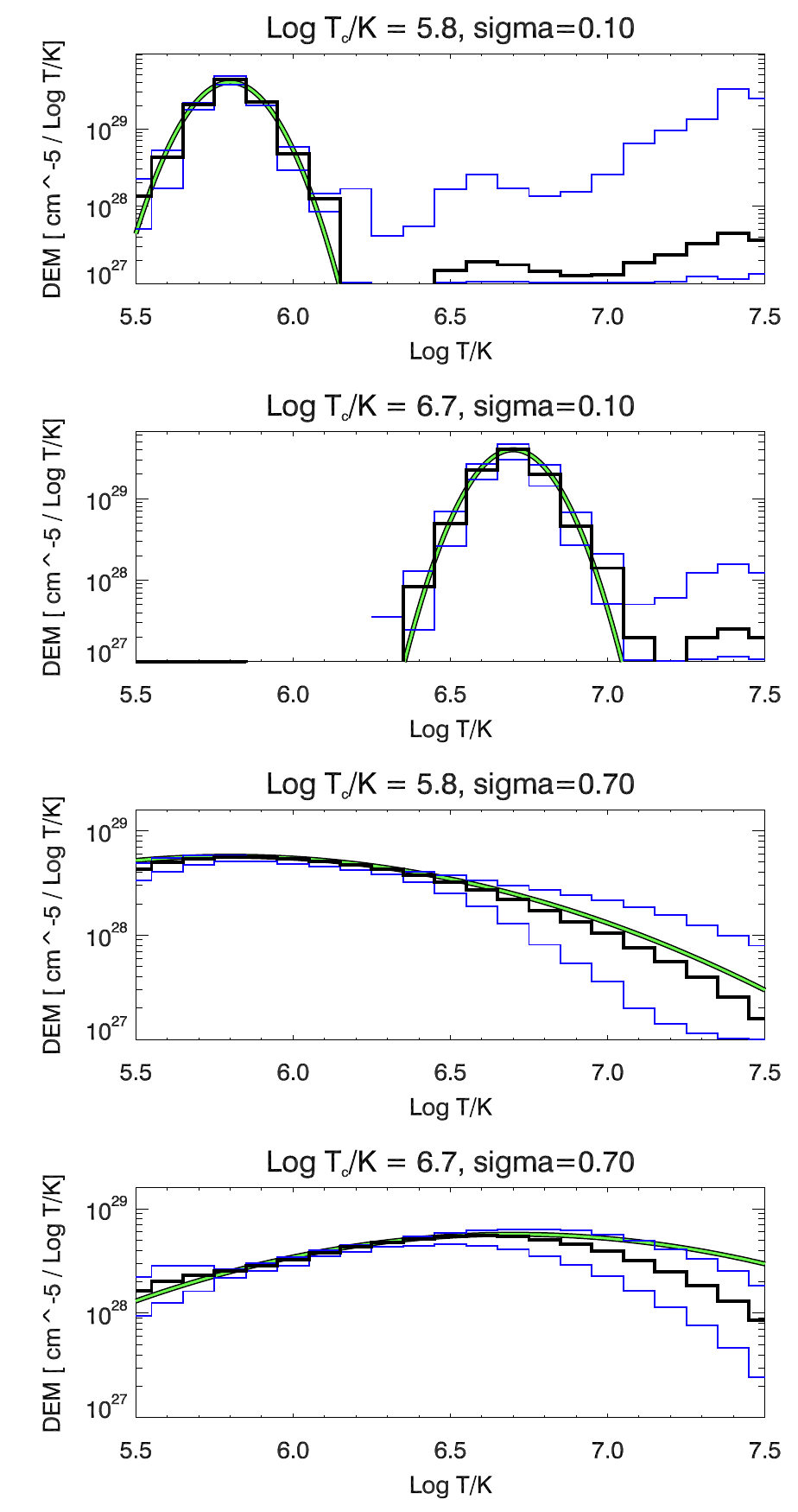}
\caption{Comparison between input Gaussian DEMs and inverted DEMs from the sparse method for four combinations of $T_c$ and $\sigma$. In each panel, the green curve indicates the DEM of the underlying model. The black solid line indicates the ensemble average of DEM solutions (for different realizations of noise). The pair of blue lines indicates the $95\%$ confidence interval, i.e. $95\%$ of the DEM solutions in the ensemble lie within the area bounded by the blue lines.}\label{fig:gaussian_curves}
\end{figure}

Inspection of Figs.~\ref{fig:gaussian_test} and~\ref{fig:gaussian_jpdfs} indicates that the sparse inversion method does not suffer from the same systematic effects encountered by~\citet{Guennou:AIADEMsII}. Even for broad DEMs ($\sigma \sim 0.7 \log T/K$), the inversion is able to return solutions that have, on average, comparable width and peak temperatures as the model inputs. Fig.~\ref{fig:gaussian_curves} shows DEMs inversions for four Gaussian models that are warm and hot ($\log T_c / K = 5.8$ and $\log T_c/K = 6.6$, respectively) and narrow and broad ($\sigma = 0.1 \log T/K$ and $\sigma = 0.7 \log T/K$, respectively). In each panel, the green curve indicates the model Gaussian DEM and the black line indicates the ensemble average of the DEM solutions. The region bounded by the pair of blue lines represents the $95\%$ confidence internal, i.e. $95\%$ of the DEM solutions in the ensemble lie within this region. The ensemble average gives a good representation of the underlying input DEM. However, some specific solutions in the ensemble do contain artifacts. For instance, in a Gaussian test case with $\log T_c = 5.8, \sigma = 0.1$ (see top panel of Fig.~\ref{fig:gaussian_curves}), some DEM solutions contain a secondary peak at $\log T/ K \gtrsim 7.0$.

There is one systematic artifact that is clearly visible in the bottom panel of Fig.~\ref{fig:gaussian_curves}. The underlying model for this test case is both hot ($\log T_c / K = 6.7$) and broad ($\sigma = 0.7\log T/K$). At $\log T / K \gtrsim 6.6$, the ensemble average has consistently lower emission measure than the underlying model. This effect can also be seen in the top right panel of Fig.~\ref{fig:gaussian_test}. For broad DEMs at high temperatures, the inversion is consistently underestimating the total EM. The middle right panel of  Fig.~\ref{fig:gaussian_test} shows that, as a result, the inversion also systematically underestimates $T_{\rm EM}$ when the underlying DEM is hot and broad. As the bottom panel of Fig.~\ref{fig:gaussian_curves} illustrates, the discrepancy is mainly due to missing EM in the high temperature bins. This is consistent with the results of~\citet{Testa:DEMReliability}, who also found the temperature systematically underestimated for broad DEMs. In section~\ref{sec:aia_xrt}, we investigate whether using both AIA and XRT observations (the latter being more sensitive to high temperature plasma) helps better constrain the DEM inversions.

\subsection{Nonlinear force-free AR with quasi-steady loops}
\label{sec:nlfff}
In the second validation exercise, we use DEM distributions from 3D thermal models of NOAA
AR 11158 by Malanushenko \& Schrijver (2015, in preparation). Construction of 3D thermal models of this AR begins with a non-linear force-free magnetic field constructed to match observed AIA loop features~\citep[from][]{Malanushenko:AR11158}. The space-filling force-free field is then decomposed into a large number of thin flux tubes (over 7000). Flux tubes are defined as the volumes enclosed between adjacent field lines, which are traced from a regular grid of seed positions. 

\begin{figure*}
\includegraphics[width=\textwidth]{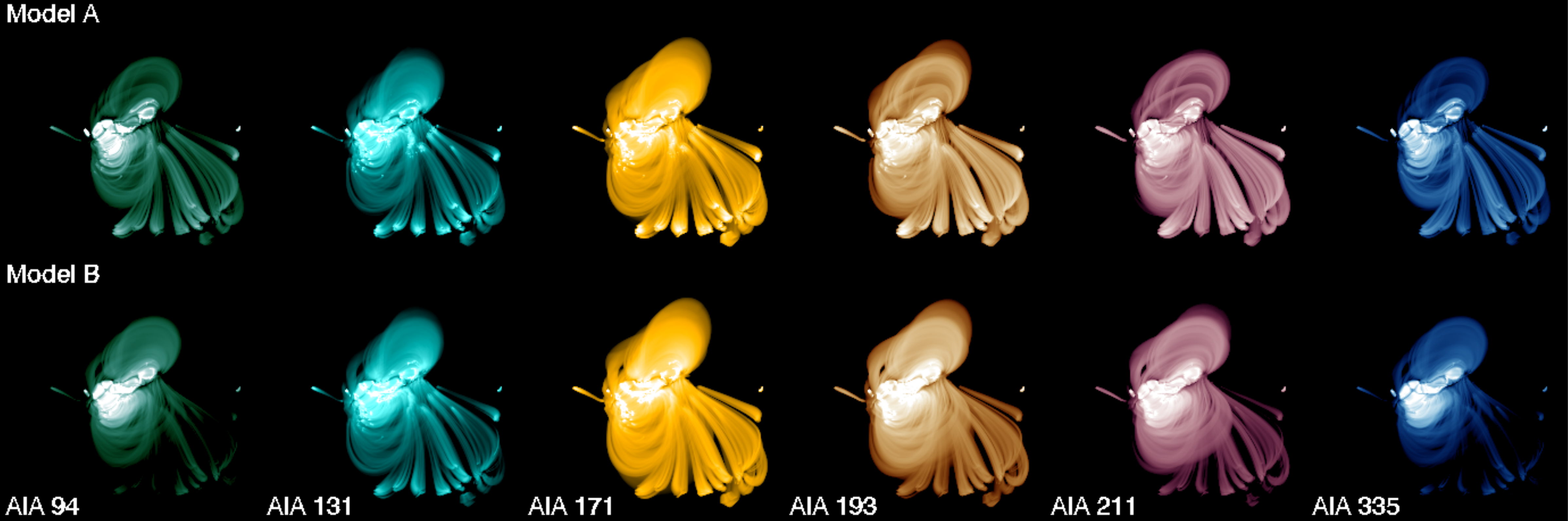}
\caption{Synthetic AIA images (log-scaled) for the two thermal models (top and bottom rows) of NOAA AR 11158.}\label{fig:anny_aia_images}
\end{figure*}

The emission of each flux tube is computed individually assuming a 1D quasi-steady
atmosphere~\citep{Schrijver:QuietSunForceFree}. In each model, the same heating scenario is used 
for the loop atmospheres of all flux tubes. The rendering method used is unique in that it
does \textit{not} assume circular cross-sections of flux tubes. Instead, the method takes into account
the distortions experienced by the flux tubes along their arc lengths~\citep{Malanushenko:Anisotropy}. \revagain{Given the set of loops for a model, the loop temperatures and densities were then resampled onto a Cartesian grid with $2.4$ arcsec pixel separation (corresponding to 4 pixels for AIA at disk center). Integration along vertical columns was performed to compute DEM distributions for all pixels. The resulting DEM data cube was then folded with AIA response functions to obtain the synthetic AIA images shown in Fig.~\ref{fig:anny_aia_images}. Although the pixel separation of the synthetic images corresponds to 4 AIA pixels, no pixel summing has been performed for the synthetic count rates (so this is akin to sampling AIA full resolution images at every $4$-th pixel in both spatial dimensions).} 

The heating scenarios for the data used here are set as follows. The volumetric heating rate $\epsilon(s)$ along each
flux tube is $\epsilon(s)\propto\Phi L^{-2.5} B (s)^{-0.5}$, where $\Phi$ is the magnetic enclosed within a tube,
$L$ is the length of the flux tube, $s$ is the arc length coordinate and $B(s)$ is the local magnetic field strength. The
proportionality constant is given in terms of the total heating energy input in the AR,
$\sum{E_H}=5\times 10^{32}$ erg s$^{-1}$, where the sum is taken over all flux tubes. The energy flux entering each flux tube is a function of base field strengths, given by
\begin{equation}
E_H\propto L^{-1.5}\Phi\left[B_1^{\beta-1}f(B_1)+B_2^{\beta-1}f(B_2)\right].\label{eqn:anny_heating}
\end{equation}
\noindent Following~\citet{Schrijver:CoronalHeating}, we set $f(B_{\rm base})=\exp \{-(B_{\rm base}/500{\rm~G})^2\}$.

\begin{figure*}
\includegraphics[width=\textwidth]{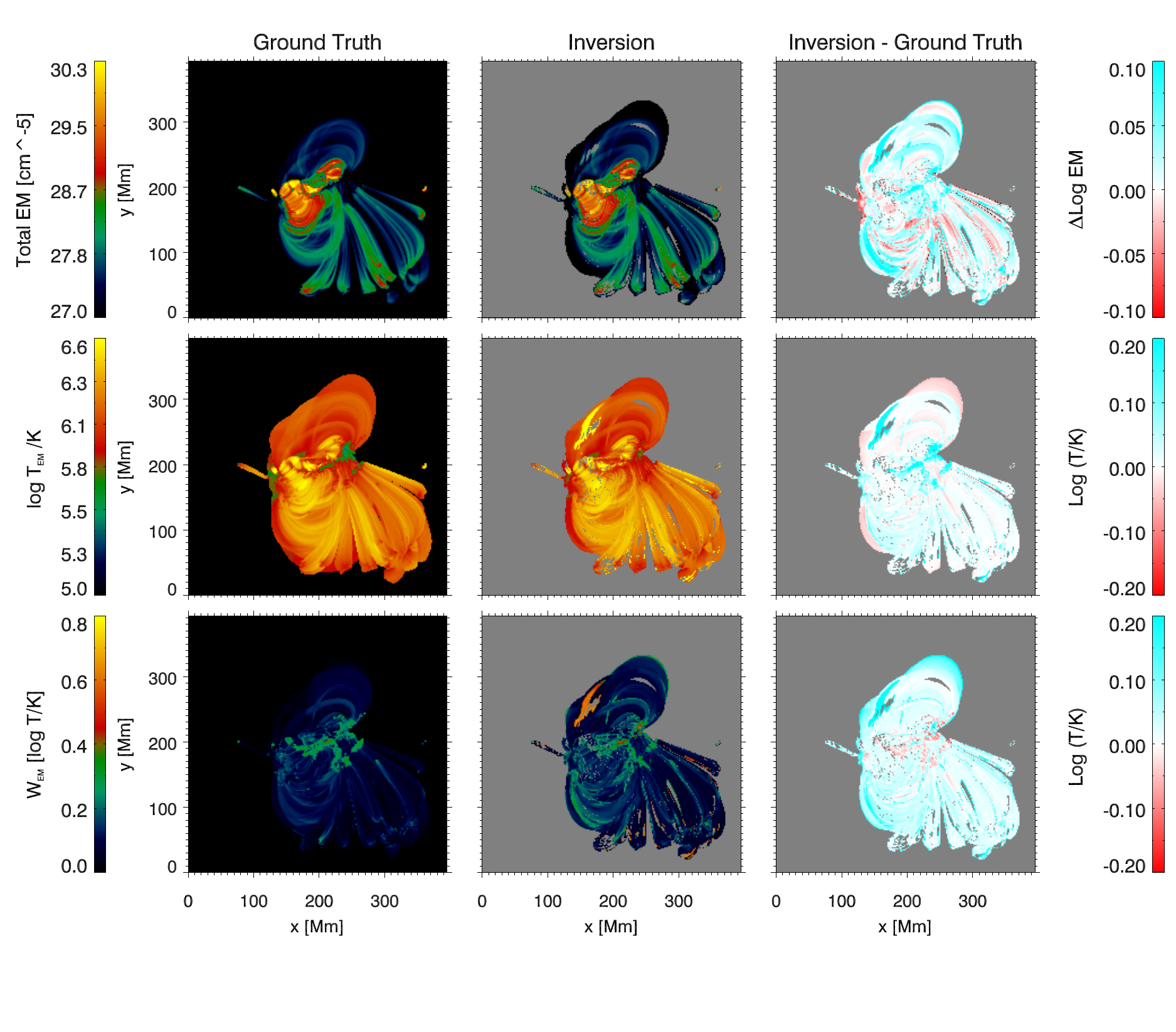}
\caption{Validation test on DEM distributions from model A ($\beta=0$ in Eq.~(\ref{eqn:anny_heating}) ) of AR 11158. The layout is the same as Fig.~\ref{fig:gaussian_test}, except the $x$- and $y$-axes here indicate spatial location in the MHD model.~\revagain{Grey pixels indicate regions where the total EM $<10^{26}$ cm$^{-5}$ and the synthetic counts are too low for DEM inversions (entire area outside of AR) or where the inversion finds no solutions (isolated pixels with the AR)}. The same comparison for a different thermal model ($\beta=2$) is shown in Fig.~\ref{fig:anny_set2}.}\label{fig:anny_set1}
\end{figure*}

\begin{figure*}
\includegraphics[width=\textwidth]{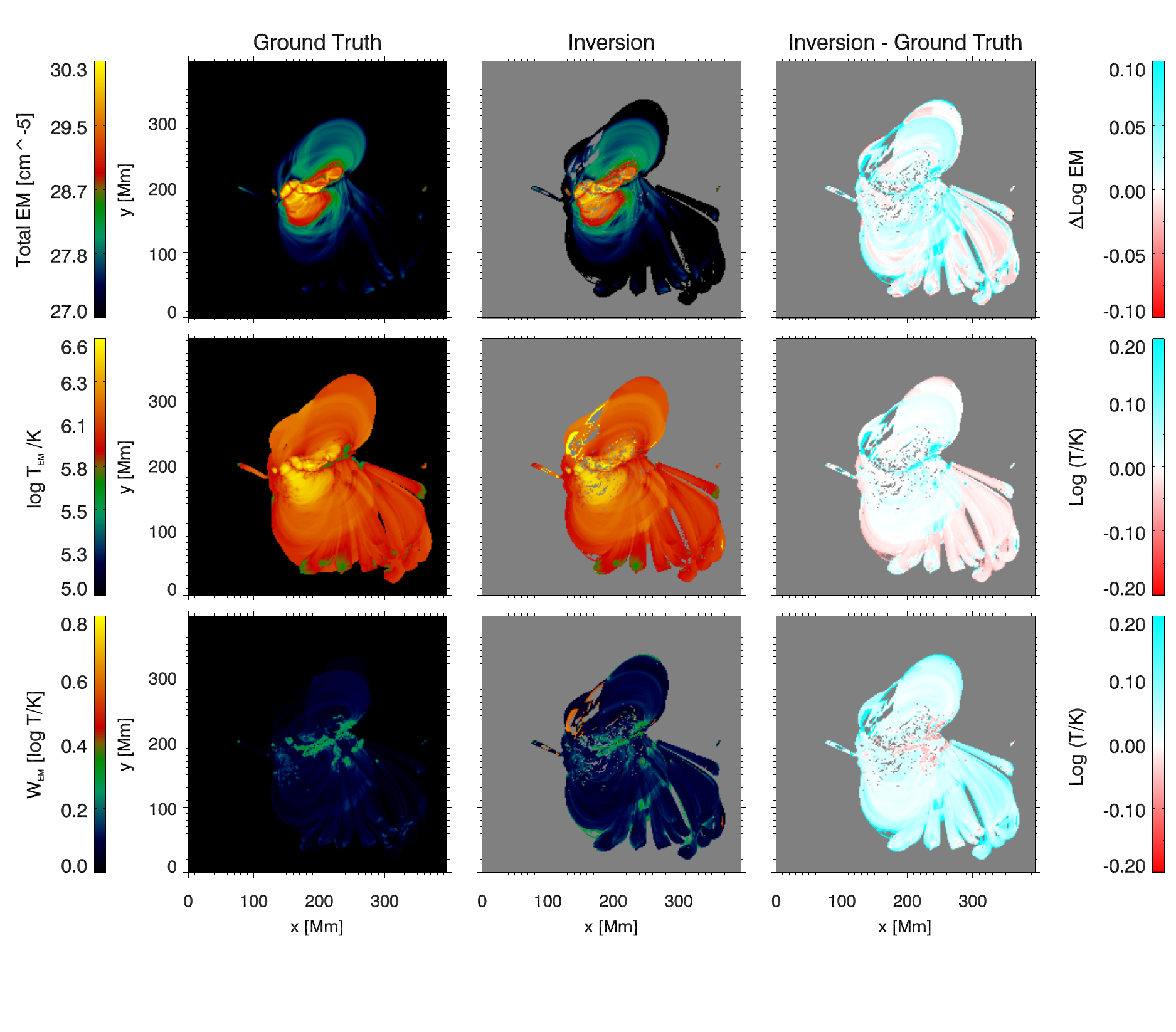}
\caption{Validation test on DEM distributions from model B ($\beta=2$ in Eq.~(\ref{eqn:anny_heating}) ) of AR 11158.  \revagain{The layout is the same as in Fig.~\ref{fig:anny_set1}, which shows a similar comparison for a different thermal model ($\beta=0$) of the AR.}}\label{fig:anny_set2}
\end{figure*}

 We consider two models for our validation exercise. Model A is computed with $\beta=0$ and model B is computed with $\beta=2$. Fig.~\ref{fig:anny_aia_images} shows synthetic AIA images for the two thermal models. These were used as input for the sparse inversion code using the same choice of a $\log T$ grid as for validation test on the log-normal distributions. \revagain{The settings for the inversion used here and in section~\ref{sec:mhd} are the same as for section~\ref{sec:gaussian_test}. However we do not perform ensemble studies of each pixel as in section~\ref{sec:gaussian_test} and we do not add noise to the synthetic data.} \revision{Rather the purpose is to see how the inversion method performs against a diverse set of DEM shapes resulting from different physical assumptions (in contrast to the idealized log-normal forms considered in the previous section). } Figures~\ref{fig:anny_set1}~and~\ref{fig:anny_set2} show the comparison between the model and inversion for models A and B, respectively. \revagain{Regions where the comparison is not performed are indicated in grey. These are places where the total EM is very low ($<10^{26}$ cm$^{-5}$, too low for inversions given AIA's sensivitity; entire grey region outside of the AR) or where the the inversion finds no solutions (isolated pixels within the AR).}  For both cases, the inversion is able to retrieve values for EM and $\log T_{\rm EM}$ that match well with the underlying model. As was the case for the validation exercise using log-normal distributions, the thermal width $W_{\rm EM}$ computed from the inversion is less reliable. Nevertheless, what is clear from this comparison is that the sparse inversion code is able to distinguish the different thermal distributions underlying the two models.

\subsection{MHD simulation of Corona Formation in an Emerging Active Region}
\label{sec:mhd}

\begin{figure*}
\includegraphics[width=\textwidth]{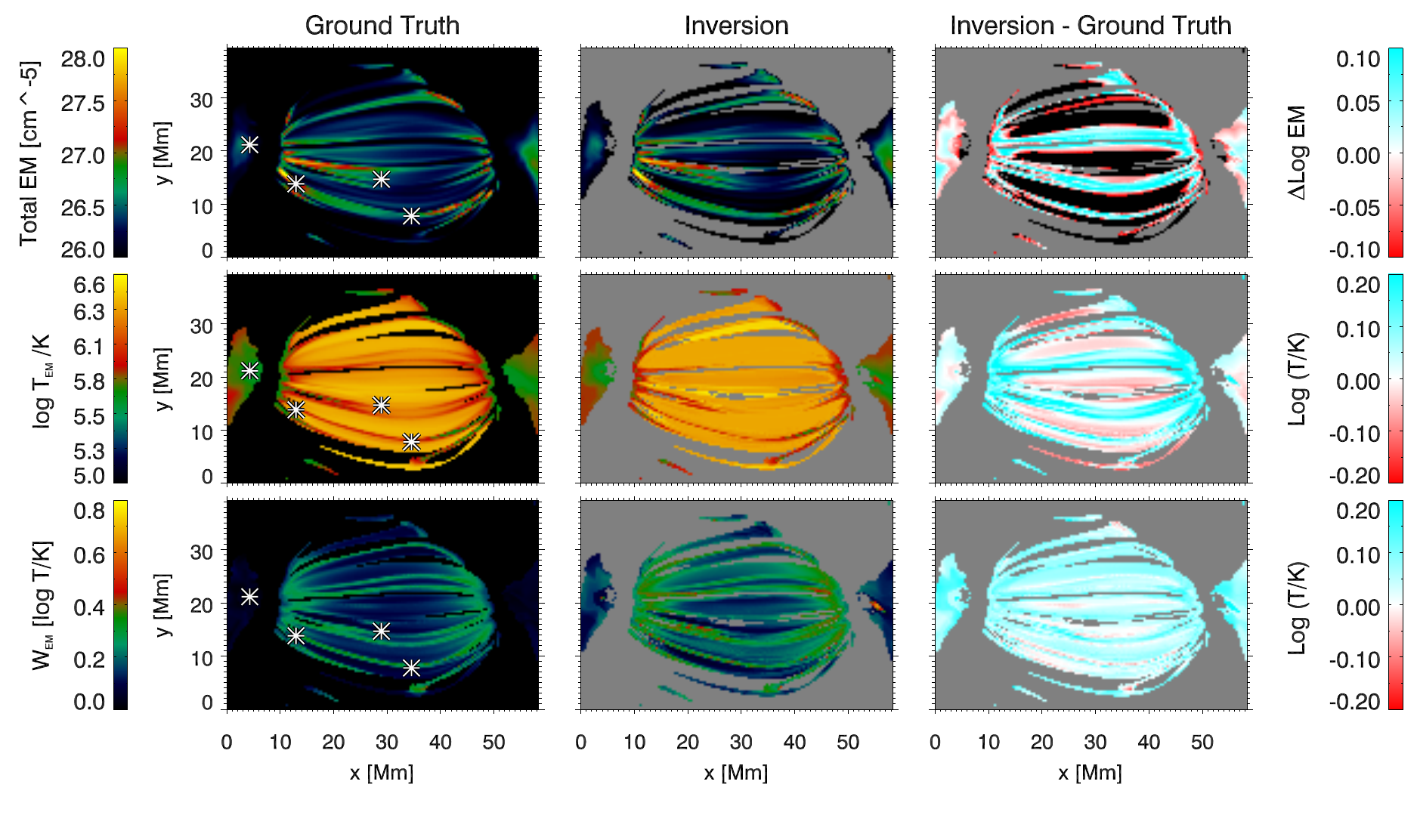}
\caption{Validation test on DEM distributions from an MHD simulation of the formation of coronal loops in an AR~\citep[][]{Chen:ARFormation}. \revagain{The layout is the same as in Figs.~\ref{fig:anny_set1} and~\ref{fig:anny_set2}. Within the AR core loops there are regions where the relative error in total EM ($\Delta \log $EM) can be of order unity (black pixels in the top right panel). This is mostly restricted to regions where total EM $\lesssim 10^{27}$ cm$^{-5}$.} The same comparison for a side view is shown in Fig.~\ref{fig:230_losy_binning3_compare}. DEM profiles for positions marked by asterisks are shown in Fig.~\ref{fig:fig_feng_select_DEMs}.}\label{fig:230_losz_binning3_compare}
\end{figure*}

\begin{figure*}
\includegraphics[width=\textwidth]{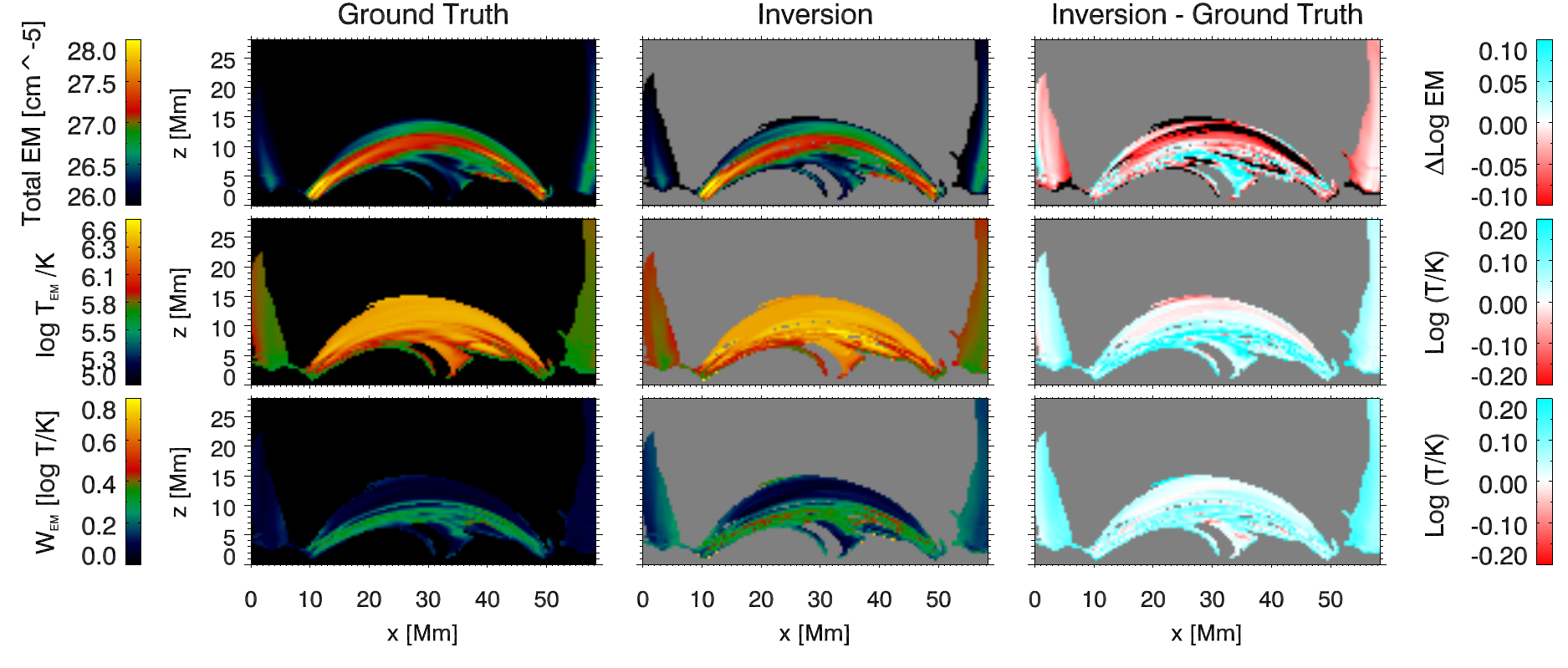}
\caption{Validation test on DEM distributions from an MHD simulation of the formation of coronal loops in an AR~\citep[][]{Chen:ARFormation}. The same comparison for a top-down view is shown in Fig.~\ref{fig:230_losz_binning3_compare}.}\label{fig:230_losy_binning3_compare}
\end{figure*}

In the third validation exercise, we use DEM distributions from a time-dependent MHD simulation of the formation of AR corona. The simulation setup is similar to the one described in~\citet[][]{Chen:ARFormation} except the numerical grid spacing is finer. The fully-compressible MHD simulation was performed using the~\emph{Pencil} code ~\citep{BrandenburgDobler:MHDTurbulence,BingertPeter:IntermittentHeating} and includes a realistic treatment of magnetic field aligned thermal conduction in the solar corona. The cartesian domain of the simulation spans $147.5\times 73.7$ Mm$^2$ in the horizontal directions. The bottom boundary is located at $z=0$ Mm (base of the photosphere) and the top boundary is at $z=50$ Mm. At the top boundary the magnetic field is matched to a potential field and the mass and thermal conductive fluxes are set to zero. Periodic boundary conditions were imposed for the side boundaries. 

The bottom boundary of the simulation was driven by imposing MHD quantities sampled from an AR formation simulation described in~\citet{RempelCheung:ARLifeCycle}. The latter captures the passage of an untwisted toroidal flux rope through the upper $15.5$ Mm of the convection zone and its eventual emergence into the overlying photosphere. However the~\citet{RempelCheung:ARLifeCycle} model has a computation domain that stops at a few hundred km above the photospheric base and so does not capture the corona. By using MHD quantities at the photosphere to set the bottom boundary condition for coronal simulation, the model of~\citet{Chen:ARFormation} demonstrated how hot coronal loops (at a few million K) spontaneously form following enhanced Poynting flux injection at their photospheric footpoints.

\begin{figure}
\centering
\includegraphics[width=0.49\textwidth]{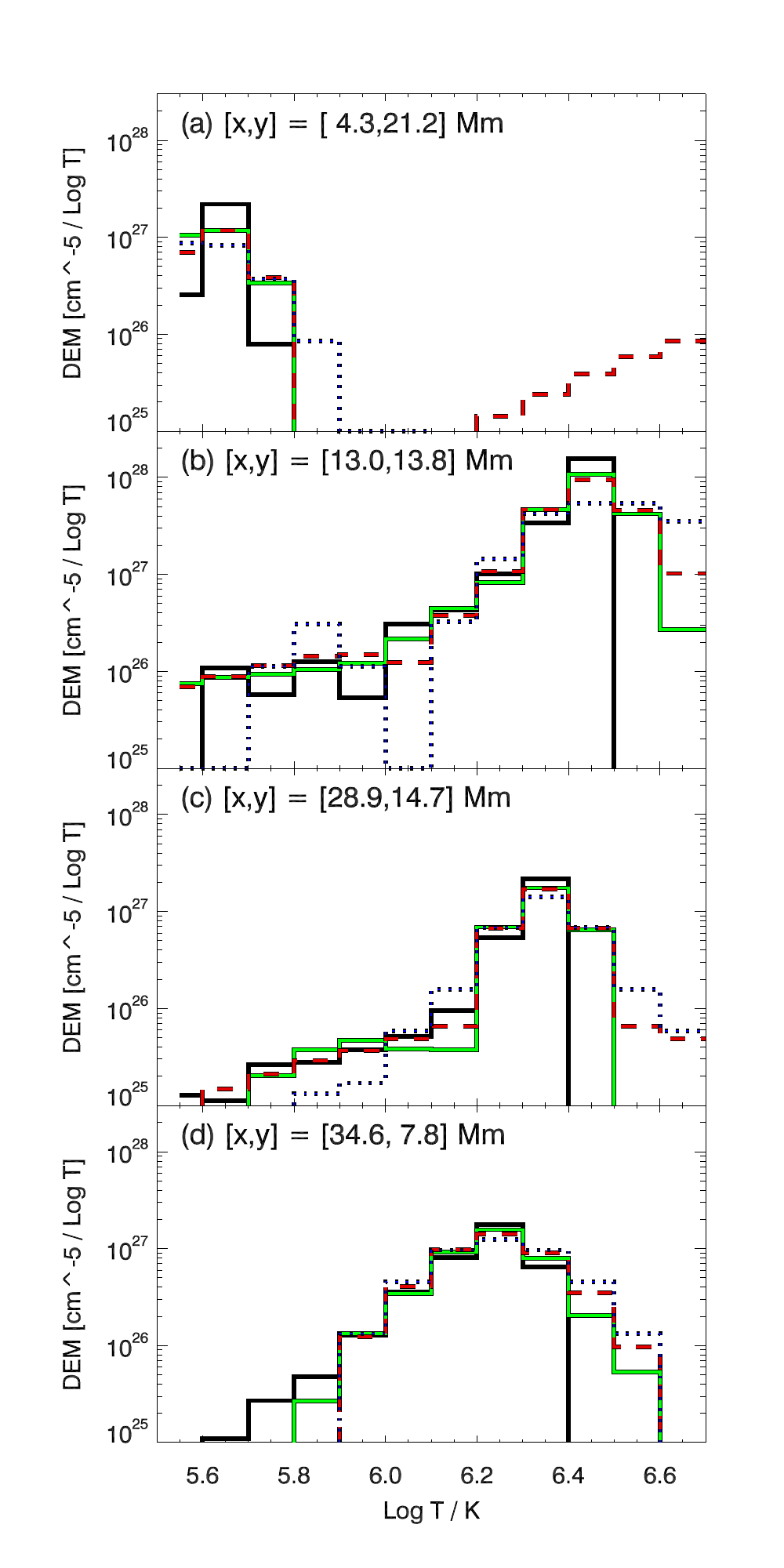}
\caption{\mcmc{Examples of DEM profiles sampled from the MHD simulation at the four positions marked by asterisks in Fig.~\ref{fig:230_losz_binning3_compare}. The black solid lines indicate the underlying DEMs from the simulation (i.e. ground truth).} \revision{In each example, three solutions are shown, corresponding to inversions with tolerance levels with no pixel averaging (blue dotted lines), and for averaging over $9$ (red dashed lines) and $36$ pixels (green solid lines).}}
\label{fig:fig_feng_select_DEMs}
\end{figure}

\revagain{For this validation exercise, DEMs were computed by sampling the simulation cubes of temperature and density (in a Cartesian domain). Line-of-sight integration of the DEM cubes was then performed for a top-down and a side view. In both cases, the spatial extent of each integration column is $432\times 432$ km$^2$. This is approximately equal to the size of an AIA pixel at disk center. The DEM maps were then used for forward synthesis of the six AIA EUV images ($94$, $131$, $171$, $193$, $211$ and $335$), which where then used as input for the sparse inversion. As in section~\ref{sec:nlfff}, the primary purpose of the exercise here is to test the inversion method against input DEMs of different shapes resulting from different physical models. Here we do not test the inversions on ensembles of noisy synthetic data as we did in section~\ref{sec:gaussian_test}. Figure~\ref{fig:230_losz_binning3_compare} shows a comparison of total EM, $\log T_{\rm}$ and $W_{\rm EM}$ between the simulation and inversion results for a top-down view. Fig.~\ref{fig:230_losy_binning3_compare} shows the corresponding comparison for a side view. Pixels with total EM $<10^{26}$ cm$^{-5}$ (in the simulation) are omitted from the comparison.} The comparisons indicate that the sparse DEM method is able to retrieve meaningful information about the thermal structure of the active region. In both the top-down and side views, the dominant feature is a set of closed loops at $\log T_{\rm EM}/K \sim 6.3$. This set of loops connect the inner edges of opposite polarity spots in the model AR. Loops that emanate from the outer edges of the spots are less inclined from the vertical, fan away from the interior of the AR and have $\log T_{\rm EM}/ K \sim 5.9$. This is near the peak of the temperature response function of the $171$~\AA~channel in AIA (see Fig.~\ref{fig:response}). The presence of these cool peripheral loops in the model (and in the inversion solution) is consistent with actual $171$~\AA~images of active regions, which often show bright fans anchored at the edge of sunspots fanning away from the interior of the AR. An example of such types of structures can be seen in Fig.~\ref{fig:AR11158}, where we applied the inversion code to actual AIA observations of an AR.

\mcmc{Figure~\ref{fig:fig_feng_select_DEMs} shows four examples of DEM profiles sampled from the MHD simulation. The sampling positions are shown as asterisks in Fig.~\ref{fig:230_losz_binning3_compare}. These profiles were selected because they have different shapes representing a diverse range of plasma conditions in the MHD model. For instance, the relatively narrow DEM profile in Fig.~\ref{fig:fig_feng_select_DEMs}(a) is sampled from a cool fan loop with a peak at $\log T/K \approx 5.75$. In contrast, the DEM in Fig.~\ref{fig:fig_feng_select_DEMs}(b) is sampled near the footpoint of a core AR loop with a peak at $\log T /K  = 6.45$ accompanied by a broad tail in the DEM distribution at lower temperatures. The profile shown in  Fig.~\ref{fig:fig_feng_select_DEMs}(c) also has a peak at $\log T /K  = 6.45$ followed by a broad tail. 
Finally, the profile shown in Fig.~\ref{fig:fig_feng_select_DEMs}(d) has a single peak at $\log T/K  = 6.25$.}

Recall that the sparse inversion method solves the linear program LP1 as given by Eq.~(\ref{eq:e3}). \revision{The solution maps shown in Figs.~\ref{fig:230_losz_binning3_compare}~and~\ref{fig:230_losy_binning3_compare} were inverted using uncertainty estimates computed assuming no averaging was performed on AIA EUV pixel values. The effect of pixel averaging (either spatially or temporally) on the inversion results can be illustrated by computing the effective $e_i$'s consistent with an averaging operation (\texttt{aia\_bp\_estimate\_error} supports this) and then using those values for the tolerance function of the inversion. This effect is illustrated in Fig.~\ref{fig:fig_feng_select_DEMs}, which shows inversion solutions for three cases: (1) no pixel averaging (dotted blue lines),  (2)  averaging over $9$ (red dashed lines) pixels, and (3) averaging over $36$ (green solid lines).}

\revision{The quality of the inversions depends on the signal-to-noise ratio of the data under consideration. For this reason, one needs to be cautious in order not to over-interpret inversion solutions. For instance, for each of the cases shown in Fig.~\ref{fig:fig_feng_select_DEMs}, the detailed shape (especially in the high temperature bins) of the DEM solution is sensitive to the noise level.} This implies the details of the DEM solution may not be very well constrained (especially in noisy data). However, it is also clear that the inversion can clearly distinguish different types of DEM profiles. Whether a DEM inversion (and its solutions) is sufficiently good depends on the science question. If the aim were to accurately measure slopes of DEM curves in the high temperature range (e.g. for constraining coronal heating models), inversions with AIA data may not be the right approach~\citep[][also find that slopes of DEMs in the high temperature bins have large uncertainties]{Warren:HighTEmissionInARs}.  However, as demonstrated in this paper, DEM inversions using AIA data can be sufficient for distinguishing the gross thermal properties between different types of coronal environments (e.g. fan loops vs. AR core loops).

\section{Application to AIA data}
\label{sec:aia_data}

The validation exercises in the previous section suggest the sparse DEM inversion method is a potentially powerful tool for helping us understand the thermal structure and evolution of coronal plasma. In this section we proceed to apply the method to actual AIA observations to illustrate its utility.

\begin{figure}
\centering
\includegraphics[width=0.5\textwidth]{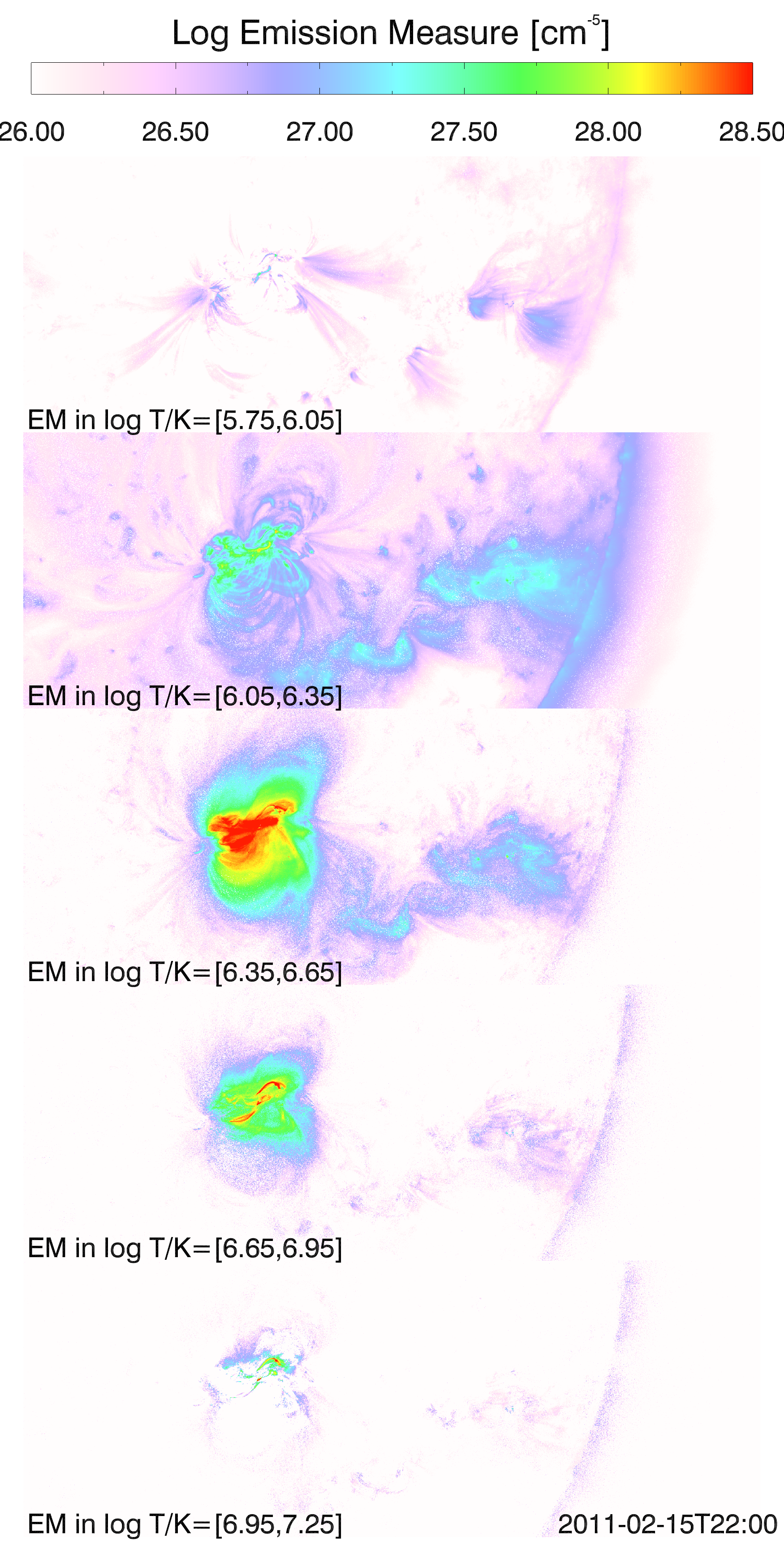}
\caption{DEM maps of NOAA AR 11158 and its surroundings obtained by the sparse inversion method. The six AIA EUV images were taken at~\texttt{2011-02-15T22:00}. The field-of-view spans $1200$$\times 480$ arcsec$^2$ and is centered at $(x,y) = (600,-268)$ arcsec.  The color-coding indicates the total EM contained within a $\log T$ range indicated in the bottom left corner of each panel.}\label{fig:AR11158}
\end{figure}

Figure~\ref{fig:AR11158} shows EM maps from a sparse inversion of AIA observations of NOAA AR 11158 and its surroundings. In the course of its appearance, this eruptive AR spawned an X$2.2$ flare~\citep[see, e.g.][]{Schrijver:AR11158,Sun:AR11158}, two M-class flares and more than a dozen C-class flares. Fig.~\ref{fig:AR11158} shows the AR at~\texttt{2011-02-15T22:00}. This is more than $20$ hours after the peak time of the GOES soft X-ray flux during the X-flare event. However, AR 11158 continued producing a series of C-class flares well after the X-flare. The most recent flare that occurred prior to~\texttt{2011-02-15T22:00} was a C6.6 flare, which began at~\texttt{2011-02-15T19:30} and ended at~\texttt{2011-02-15T20:53}.

The top panel of Fig.~\ref{fig:AR11158} shows the total EM contained within the temperature range $\log T \in [5.75,6.05]$. One can clearly identify fan loops anchored at the east and west edges of the AR. These cool fan loops are generally oriented away from the AR core~\citep[see also][]{Brooks:FanLoops}.   In the temperature range $\log T\in[6.05,6.65]$ (the next two panels), the high-EM areas mostly delineate core loops connecting the leading and following polarities of the AR. This transition from cooler peripheral fan loops to warmer core loops is consistent with the trend we saw in the MHD simulation discussed in section~\ref{sec:mhd}. In this phase of its life, the AR is still flare productive and has high-EM in distinct loops even at temperatures above $\log T/K \sim 6.6$.

When we inspect the DEM maps of the same AR one month later (i.e., one solar rotation later), we find very distinct differences. At this time, the AR is well into its decay phase. Overall, the AR has lower total EM. In addition, the EM in temperature bins above $\log T/K \sim 6.6$ is diminished by at least $2-3$ orders of magnitude. Presumably, the weaker magnetic field strengths and the lack of emerging flux and/or shear flows results in less energization of the AR complex, leading to lower EMs and EM-weighted temperatures.

\begin{figure}
\centering
\includegraphics[width=0.5\textwidth]{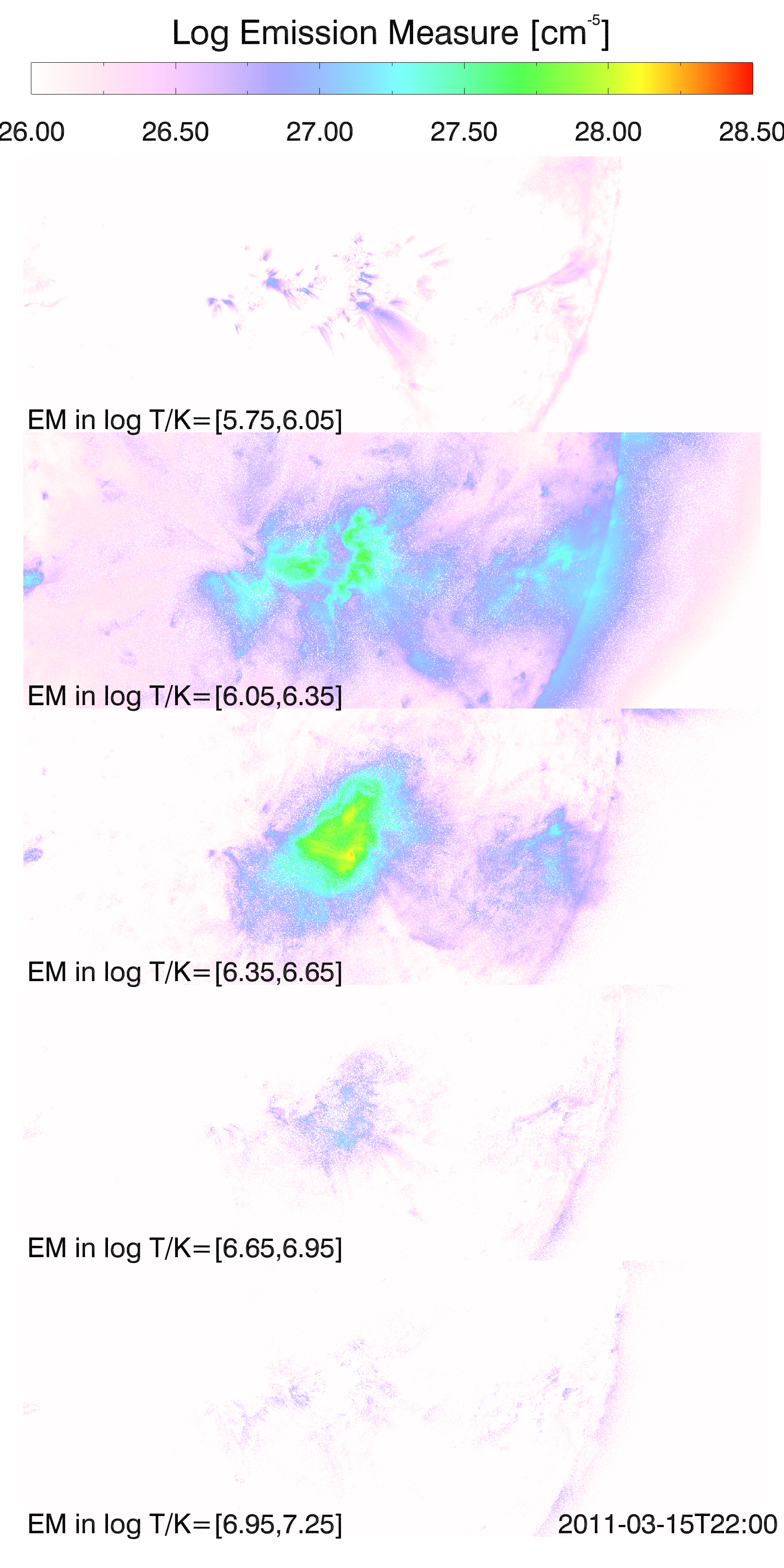}
\caption{Same as Fig.~\ref{fig:AR11158} but one month later, at ~\texttt{2011-03-15T22:00}. At this time, AR 11158 has decayed substantially and there is very little EM at temperatures above $\log T/K = 6.65$. }\label{fig:AR11158_old}
\end{figure}

\subsection{Speed}
\label{sec:speed}
The sparse inversion code is fast compared to other codes described in the literature for AIA DEMs. The following performance numbers are for a single IDL thread running on a MacBook Pro with a 2.6 GHz Intel Core i7 processor. For an inversion problem with six AIA channels ($m=6$), a temperature grid with $n=21$ bins and a set of $l=4\times 21=84$ basis functions (see Appendix~\ref{sec:quadrature}), the sparse inversion code computes more than $10^4$ solutions per second. This does not include the one-time initialization step of setting up the response matrix, which is typically a few seconds and is dominated by disk I/O performed by the Solarsoft routine~\texttt{aia\_get\_response}. However the start-up time can be amortized as the number of observation vectors increase.

\section{Joint AIA \& XRT DEM inversions}
\label{sec:aia_xrt}
In our validation tests with log-normal distributions (see section~\ref{sec:gaussian_test}), we found that the sparse method is able to return relatively reliable estimates of the total EM, EM-weighted $\log T$ and (with higher uncertainty) the thermal width of the model distribution functions. However, there was one systematic effect, namely that for hot and broad DEM distribution, the method had a tendency to underestimate the amount of EM at $\log T/K\gtrsim 6.7$ (see Fig.~\ref{fig:gaussian_curves}). This raises the question of whether the inclusion of an X-ray channel could help improve the DEM solution. So we repeated the validation exercise with one modification. Instead of just six AIA EUV channels, we augment the observation vectors $\vec{y}$ with an additional component corresponding to synthetic count rates observed by the Be-thin channel of XRT. To synthesize the XRT count rates we folded the model DEM functions with the temperature response function given by the Solarsoft routine~\texttt{make\_xrt\_temp\_resp}. This broadband X-ray channel has a response function that peaks at $\log T/K = 7.0$~\citep[][]{Golub:XRT}.

\begin{figure*}
\includegraphics[width=\textwidth]{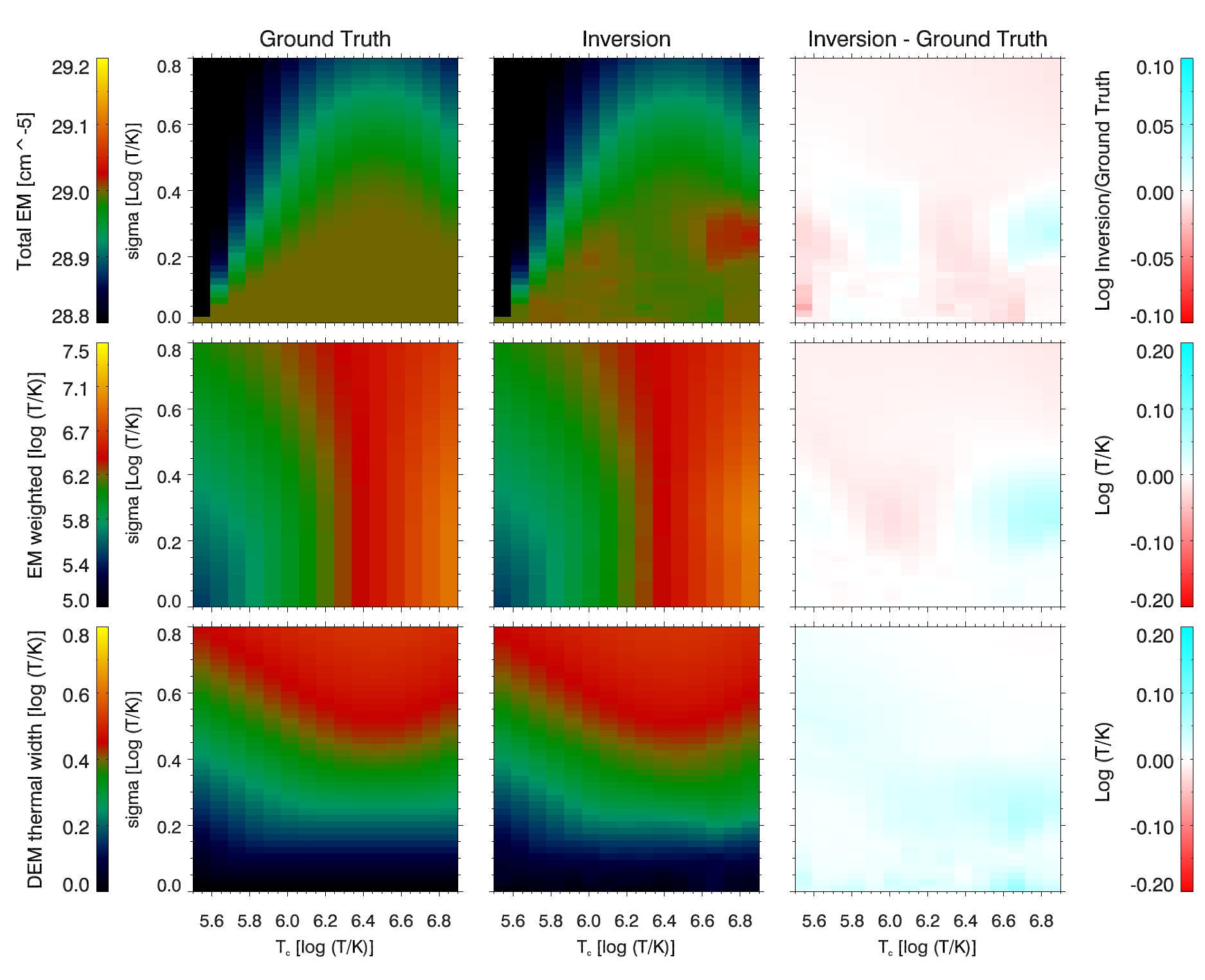}
\caption{Similar to Fig.~\ref{fig:gaussian_test} except here we performed the DEM inversions on observation vectors comprising six AIA EUV channels ($94$, $131$, $171$, $193$, $211$ and $335$) and the Be-thin channel from XRT. Noise was added to the synthetic count rates to generate an ensemble of noisy observation vectors for each model. The top, middle and bottom panels show ensemble averages of the three metrics used the quantify the fidelity of the DEM solutions (total EM, EM-weighted log T and thermal width) to the underlying DEM model.}\label{fig:gaussian_test_xrt}
\end{figure*}

\begin{figure*}
\includegraphics[width=\textwidth]{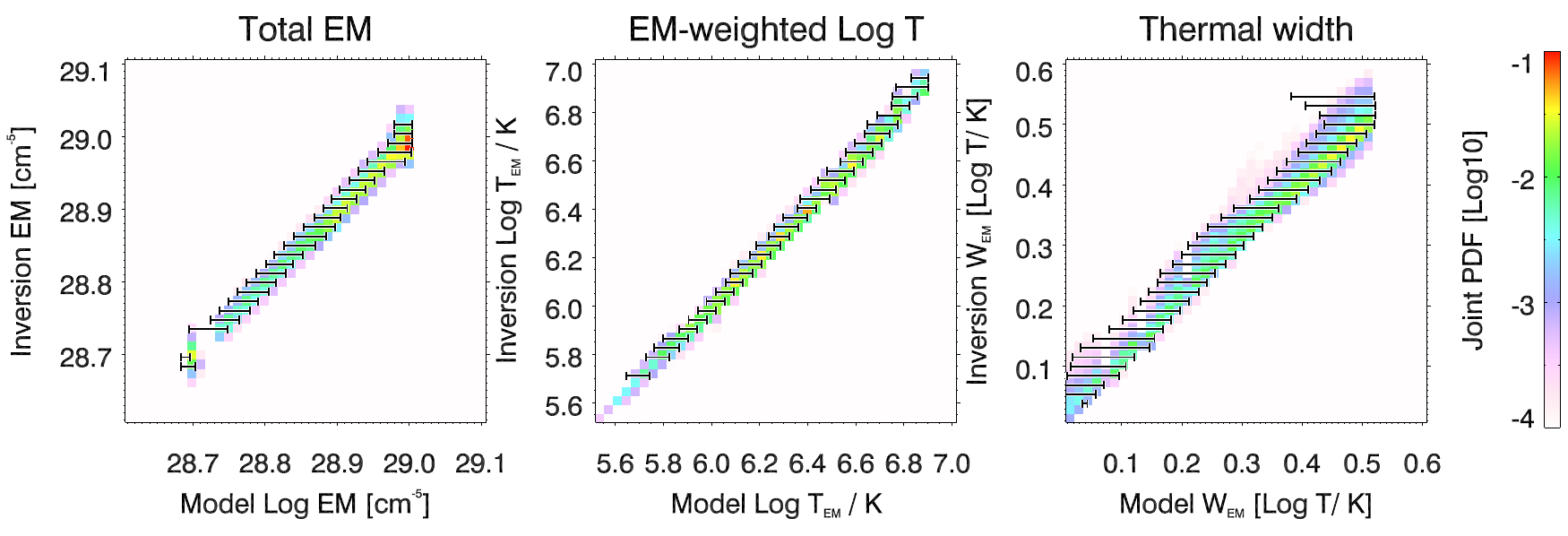}
\caption{Similar to Fig.~\ref{fig:gaussian_jpdfs} except here we performed the DEM inversions on observation vectors comprising six AIA EUV channels ($94$, $131$, $171$, $193$, $211$ and $335$) and the Be-thin channel from XRT. Compared to using AIA data only, the joint AIA-XRT inversion provides DEM solutions that better match the underlying model.}\label{fig:gaussian_xrt_jpdfs}
\end{figure*}

Figure~\ref{fig:gaussian_test_xrt} shows a comparison of $\langle {\rm EM} \rangle$, $\langle \log T_{\rm EM} \rangle $ and $\langle W_{\rm EM} \rangle$ for both the model and the inversion solutions. When we compare this figure with Fig.~\ref{fig:gaussian_test_xrt} (showing corresponding results for inversions using only six AIA channels), we find a clear improvement in the fidelity of the inversion solutions with respect to the underlying DEM distributions. The discrepancy between the inversions and the underlying model is much reduced in all three metrics. What is interesting is that this improvement is not limited to broad and hot DEMs. Even in the parameter regime where the DEMs are centered at lower temperatures (say $\log T \lesssim 6.3$), the introduction of the Be-thin channel results in a marked improvement. Fig.~\ref{fig:gaussian_xrt_jpdfs} shows the corresponding joint PDFs for the joint AIA/XRT inversions. Compared with Fig.~\ref{fig:gaussian_jpdfs}, we see how the introduction of an XRT channel tightens the $95\%$ confidence intervals for all three metrics.

This exercise demonstrates two important points. Firstly, the sparse DEM inversion method can easily be extended to take into account data from instruments other than AIA. Secondly, the inclusion of X-ray data (such as from XRT) can help improve the DEM inversion results~\citep[see also][ for a discussion of the impact of performing joint AIA-XRT DEM inversions]{Hanneman:DEMs}. In future work, we will perform joint AIA-XRT inversions on real data. This requires a careful examination of the intercalibration between the instruments and is outside the scope of the present study. 

\section{Discussion}
\label{sec:discussion}
By delivering full-disk EUV observations of the Sun at high cadence, spatial resolution and regularity, SDO/AIA has so far proven immensely valuable for studies of the dynamical behavior of coronal features in EUV images. The interpretation of optically thin EUV features in AIA images in the thermal domain requires either DEM analysis and/or sophisticated MHD modeling with forward synthesis of observables~\citep[e.g.][]{Peter:CoronalHeatingThroughBraiding,Gudiksen:AbInitioApproachCoronalLoops,Mok:ThermalStructureOfSolarARs,Hansteen:RedBlueShifts,PeterBingert:LoopCrossSections,Jin:AWSoMCME,vanderHolst:AWSoM,Chen:ARFormation}. The latter remains essential for improving one's understanding of the physical processes operating in solar plasma. Forward modeling of synthetic observables from numerical models and a direct comparison of these computed quantities with observation data provides the most direct method for testing the validity of the models. However, since numerical simulations are computationally expensive, they generally sample a very limited range of parameter regimes and physical scenarios. Thermal analysis techniques such as DEM inversions are therefore important for probing the parameter range available to solar phenomena. Using the concept of sparsity, we have implemented a high throughput DEM inversion code that returns positive semidefinite solutions with a computational speed that is suitable for producing DEM maps from AIA images.

Before applying any DEM method to actual solar data, one must validate the method against a variety of model thermal distributions to ensure the method returns solutions that are representative of the physical scenario underlying the data. To this end, we carried out validation tests~\citep[see also][]{Testa:DEMReliability,Guennou:AIADEMsI,Guennou:AIADEMsII,Plowman:FastDEMs} on three classes of DEM models. They include log-normal distributions (section \ref{sec:gaussian_test}), DEMs from quasi-steady loops loaded on non-linear force-free field models of an AR (section \ref{sec:nlfff}), and DEMs from a fully time-dependent MHD simulation of AR corona formation (section~\ref{sec:mhd}). By testing the inversion method against models of varying complexity based on different physical assumptions, we mitigate the tendency to tune the inversion method to one particular type of DEM distribution (a type of overfitting). 

To illustrate the utility of our new inversion method, we applied it on DEM inversions of AIA observations of NOAA AR 11158 at two different phases of its life. DEM maps of the AR during its flaring phase (Fig.~\ref{fig:AR11158}) show high-EM core loops above $3.5$ million K. In comparison, DEM maps of the AR during its decay phase show a clear deficit of material at similar temperatures. In both phases, high-EM structures in the low temperature bins ($0.5-1$ million K) outline fan loops predominantly anchored at the periphery of the AR.

In section~\ref{sec:aia_xrt}, we showed how supplementing AIA images with an XRT channel (e.g. Be-thin) improves the quality of sparse DEM inversions. This hypothetical example serves to illustrate how the sparse DEM method can easily be adapted and/or extended for inversion of data from other (or multi-) instruments.

\mcmc{Whether DEMs with AIA data are appropriate for probing the thermal structure of the solar corona depends on the science question. As discussed in section~\ref{sec:mhd}, the slopes of DEMs in high temperature range are not well constrained. So if this measurement were crucial for testing coronal heating models, the present approach is perhaps not appropriate. However, in this paper we have demonstrated the ability of the sparse inversion code (as applied to AIA data) to yield DEM solutions that allow one to distinguish between different types of thermal structures in the solar corona (e.g. cool fan loops vs. AR core loops).  The present work opens up the possibility of routine production of DEM maps using AIA data (from a validated method) even in the absence of complementary coverage from XRT and EIS. This allows AIA to fulfill its promise to help researchers probe the thermal distribution and evolution of the solar corona.}

\acknowledgements
\revision{The authors acknowledge support from NASA's SDO/AIA contract (NNG04EA00C) to LMSAL. AIA is an instrument onboard the Solar Dynamics Observatory, a mission for NASA's Living With a Star program. Additionally, MCMC, PT and AM acknowledge support from NASA's Heliophysics Grand Challenges Research grant (NNX14AI14G).}

\revision{The authors wish to thank members of the Hinode/XRT team (especially M. Weber and K. Reeves) for valuable discussions about joint AIA/XRT inversions. Furthermore, we wish to thank H. Warren, J. Plowman, S. White, and the anonymous referee for helpful comments that led to improvements in this work.}
\newpage
\bibliographystyle{apj}
\bibliography{references}
\appendix

\section{Quadrature scheme}
\label{sec:quadrature}
Let $i = 1,2,...,m$ denote the index over a set of wavelength band channels and/or line spectra. Let the DEM function be written in terms of a set of positive semidefinite basis functions  $\{b_j(\log T) \ge 0~|~k=1,2,...,l\}$, viz.
\begin{equation}
{\rm DEM}(\log T)= \sum_{k=1}^l b_k(\log T) x_k,
\end{equation}
\noindent with quadrature coefficients $x_k\ge 0$. Approximating the integrals in equation (\ref{eqn:yintegral}) as sums in $\log T$ space, we have 
\begin{eqnarray}
y_i = \sum_{j=1}^{n} \sum_{k=1}^l K_{ij} B_{jk }  x_k \Delta \log T,
\end{eqnarray}
\noindent where $j=1,2,...,n$ is the index over temperature bins, $K_{ij} = K_i(\log T_j)$ and $B_{jk} = b_k(\log T_j)$. The response matrix $\mathbf{K} = (K_{ij}) $ has dimensions $m\times n$. The basis matrix $\mathbf{B} = (B_{jk})$ has dimensions $n\times l$, with the $k$-th column vector corresponding to the $k$-th basis function $b_k(\log T_j)$. Defining the~\emph{dictionary  matrix} $\mathbf{D} = \mathbf{K}\mathbf{B}$, the set of integral equations (\ref{eqn:yintegral}) can be written in matrix form:
\begin{equation}
\vec{y} = \mathbf{D}\vec{x},\label{eqn:matrix_kb}
\end{equation}
\noindent where the sought-after solution vector $\vec{x}$ is an $l$-tuple with components $x_k \Delta \log T$ ($k=1,2,...,l$). When the number of basis functions exceeds the number of image channels (i.e. $l>m$), the linear system Eq. (\ref{eqn:matrix_kb}) is underdetermined. 

For the results in this paper, we use an equidistant grid in $\log T$ with $\Delta \log T = 0.1$ ranging from $\log T=5.5$ to $7.5$ (i.e. $n=21$). Over this temperature grid, the set of Dirac-delta basis functions $\{ b_k^{\rm Dirac}~|~ k=1,...,n\}$ is 
\begin{eqnarray}
b_k^{\rm Dirac}(\log T_j)  &= & 1,{\rm~if}~\log T_j = \log T_k, \\ 
& = & 0 ,{\rm~otherwise}.
\end{eqnarray}
\noindent Recall that the basis matrix $\mathbf{B}$ consists of column vectors corresponding to basis functions. So for the set of Dirac-delta functions $\mathbf{B}^{\rm Dirac}=\mathbf{I}$ (the identity matrix).

\begin{figure*}
\includegraphics[width=\textwidth]{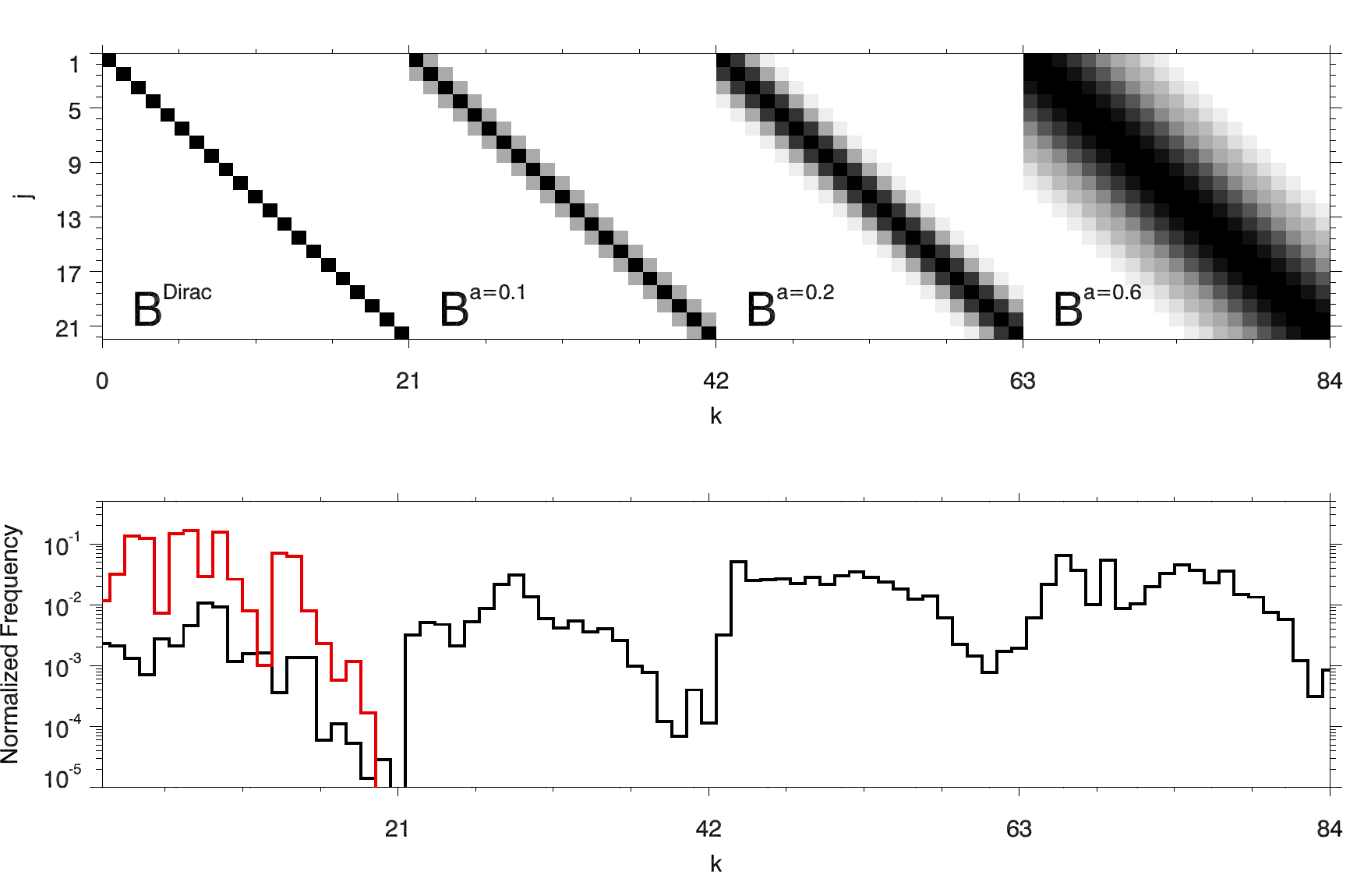}
\caption{Each column vector in the basis matrix $\mathbf{B}$ given in Eq. (\ref{eqn:basis_concat}) corresponds to a basis function. The basis matrix shown~\revagain{in the upper panel} results from four sets of basis functions. The leftmost set corresponds to Dirac-delta functions and the remaining three correspond to truncated Gaussians of width $a=0.1$, $a=0.2$ and $a=0.6 \log T/K$. In the shading used above, black indicates unity and white indicates zero. \revagain{The lower panel shows the likelihood that a basis function (i.e. a column in the basis matrix) has a non-zero coefficient in the DEM inversion test performed on log-normal DEMs (see section~\ref{sec:gaussian_test}). The black line corresponds to the case where the basis matrix shown in the upper panel (i.e. Dirac-Delta functions and three sets of Gaussians) is used. The red line corresponds to the case when only Dirac-delta functions are used as the basis.} } \label{fig:basis_funcs}
\end{figure*}

In addition to Dirac-delta functions, we also use basis functions consisting of truncated Gaussians. Each Gaussian function of width $a$ generates a set of basis functions $\{ b_k^{a}~|~ k=1,...,n\}$, where
\begin{eqnarray}
b_k^{\rm a}( \log T_j)  &=& \exp{\left[-\frac{(\log T_j - \log T_k)^2}{a^2}\right]}, {\rm~if~} |\log T_j - \log T_k |\le 1.8a.\\
& = & 0,{\rm~otherwise}.
\end{eqnarray}
\noindent \revagain{The Gaussian basis functions are truncated (i.e. set to zero) for values of $\log T_j$ outside the temperature grid used for inversions. The corresponding basis matrix for this set is denoted $\mathbf{B}^a$.} Different sets of basis functions can be combined by concatenating their associated basis matrices. For the inversions shown here, we use the combined basis matrix

\begin{equation}
\mathbf{B} = \left ( \mathbf{B}^{\rm Dirac}| \mathbf{B}^{a=0.1}| \mathbf{B}^{a=0.2} | \mathbf{B}^{a=0.6}\right).\label{eqn:basis_concat}
\end{equation}
\noindent \revagain{ Note the individual Gaussian basis functions are not normalized by their sums (i.e. all have maximum value of unity at their peaks). So given multiple solutions that equally fit the data, the method will prefer a solution consisting of a single broad Gaussian over solutions consisting of multiple narrow Gaussians (and/or Dirac-delta functions). Empirically, we find the choice of not normalizing the Gaussian basis functions results in better inversions results (more on this below). } Because $n=21$, $\mathbf{B}$ as indicated above (see Fig.~\ref{fig:basis_funcs} for a graphical representation) has dimensions $21\times 84$ and $\mathbf{D}$ has dimensions $m\times 84$. With six AIA channels, $m=6$. Even when AIA is augmented by XRT or EIS data, $m\ll 84$. This makes Eq.~(\ref{eqn:matrix_kb}) a highly underdetermined system, which we solve by the method of basis pursuit (see section~\ref{sec:sparse_method}).

Seeking to solve this underdetermined system is the same as the following geometric problem. Suppose we aim to express some given column vector $\vec{y}$ (of dimension $m$) as the linear combination of members drawn from a family of column vectors. Let this family of column vectors be denoted $\{\vec{d}_k ~|~k=1,...,l\}$. The goal is to find coefficients $x_k$ such that $\vec{y} = \sum_{k=1}^l x_k \vec{d}_k$, which is equivalent to the linear system (\ref{eqn:matrix_kb}) if the dictionary matrix $\mathbf{D}$ is constructed by concatenating the $\vec{d}_k$'s side by side. Because $l > m$, $\{\vec{d}_k ~|~k=1,...,l\}$ is an overcomplete set of possible basis vectors (i.e. dictionary) for building up $\vec{y}$. So the non-uniqueness of a DEM solution satisfying Eq. (\ref{eqn:matrix_kb}) is the same as the multiplicity of ways to find a basis for $\vec{y}$. Basis pursuit addresses this by seeking a solution that minimizes the L1-norm $|\vec{x}|_1$. In other words, basis pursuit finds the most sparse representation of $\vec{y}$ from an overcomplete dictionary $\mathbf{D}$~\citep[][]{Chen:AtomicDecomposition}.


\begin{figure*}
\includegraphics[width=\textwidth]{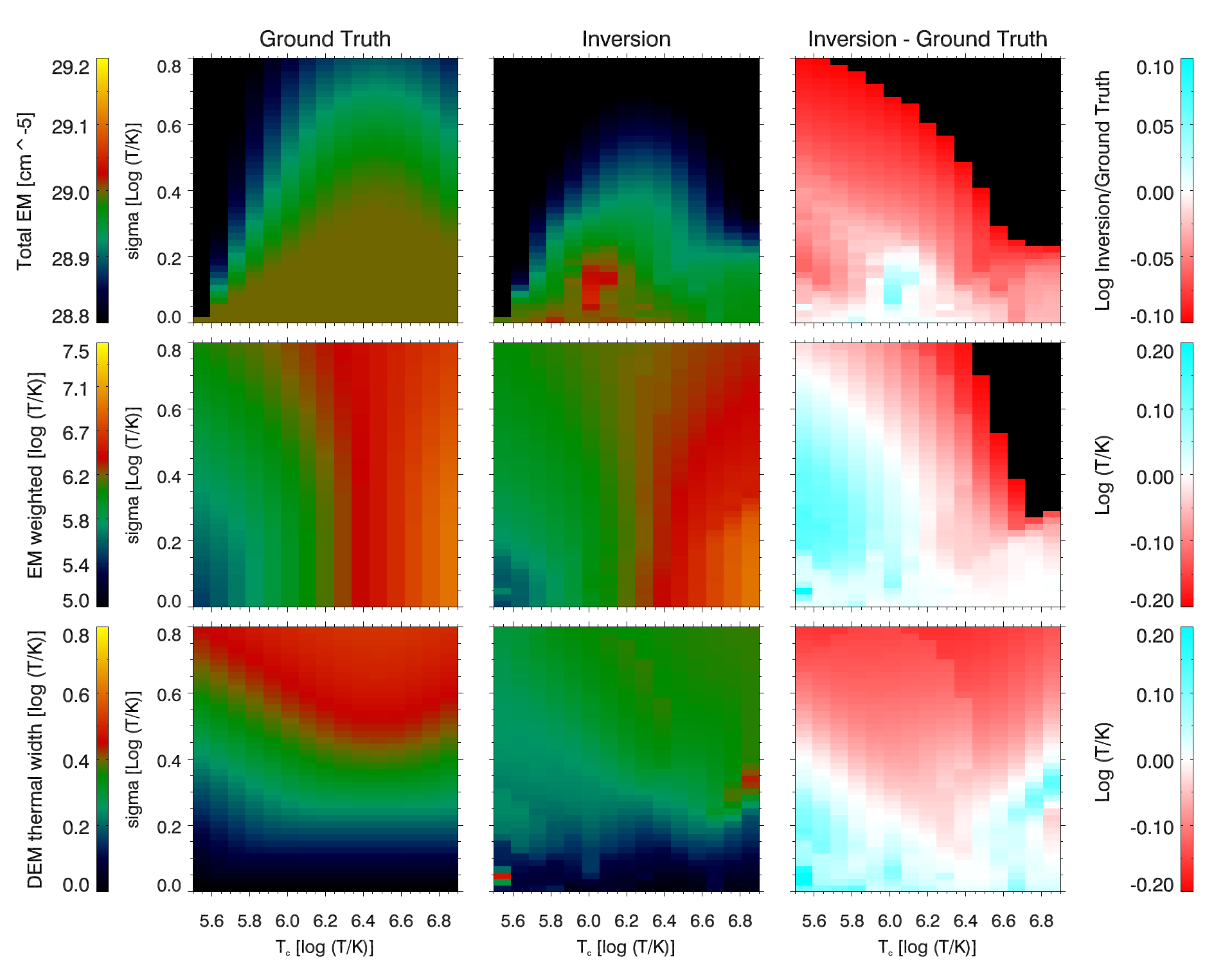}
\caption{Similar to Fig.~\ref{fig:gaussian_test} but for a quadrature scheme that uses only Dirac-delta basis functions. The quality of the DEM inversions is clearly inferior in this case. The black patches in the top right corners of the plots means the inversion is returning total EMs and EM-weighted $\log T$ values with discrepancies exceeding the range displayed by the color table.}\label{fig:gaussian_test_dirac}
\end{figure*}

\begin{figure}\centering
\includegraphics[width=0.5\textwidth]{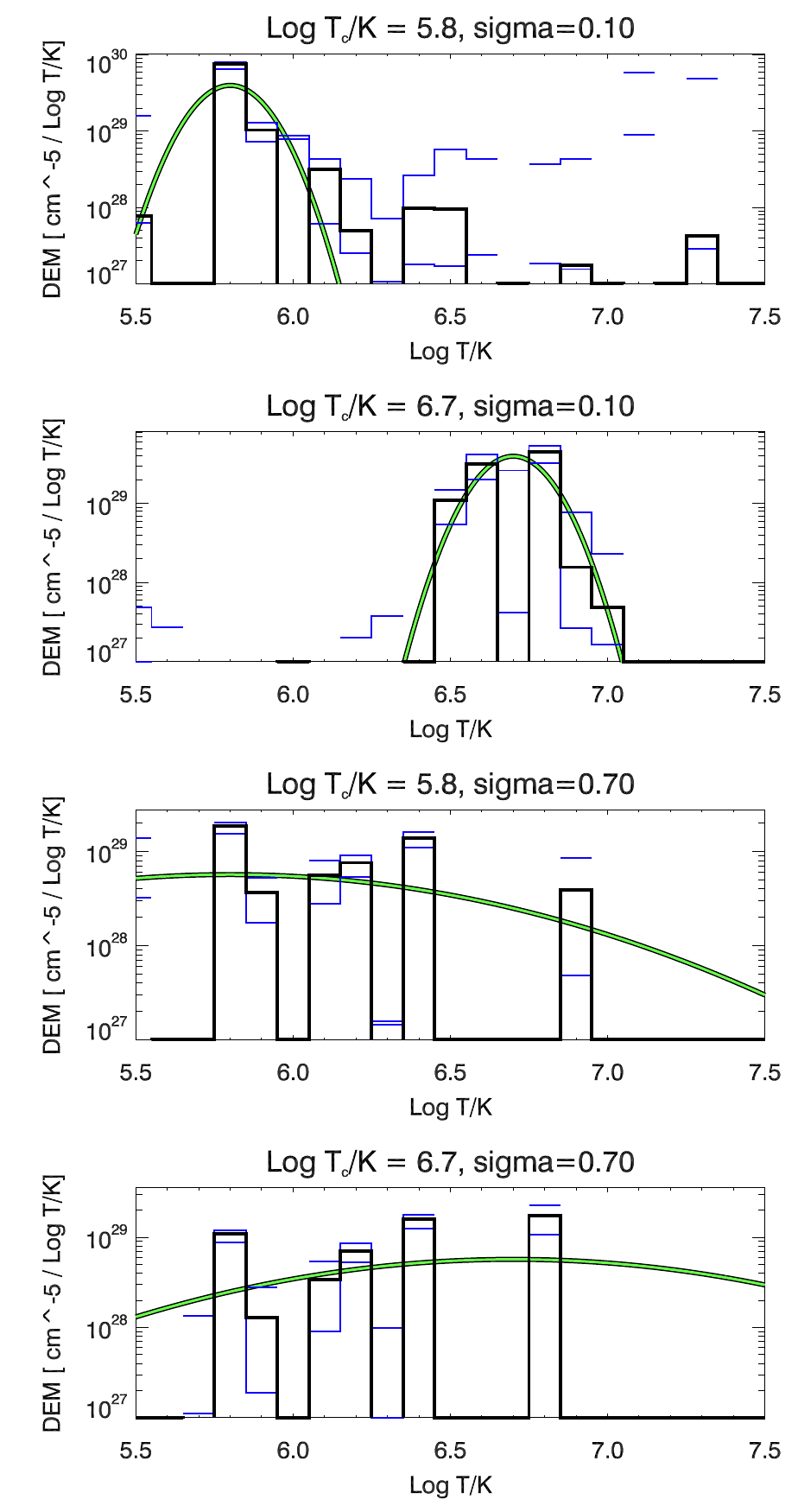}
\caption{\revagain{Similar to Fig.~\ref{fig:gaussian_curves} but for a quadrature scheme that uses only Dirac-delta basis functions. The quality of the DEM inversions is clearly inferior in this case. Here the inversion solutions are expressed as the sum of a few isothermal components and result in distributions that are much more spiked than the underlying DEM distributions.}}\label{fig:gaussian_curves_dirac}
\end{figure}

It is worth comparing the fidelity of the inversions for different bases. For the results shown throughout this paper, we used the basis given by Eq. (\ref{eqn:basis_concat}). This was a choice made after performing validation exercises for a number of different bases. For example, consider the validation exercise on log-normal DEMs (section~\ref{sec:gaussian_test}). Fig.~\ref{fig:gaussian_test} shows a comparison between the model DEMs and the inversion solutions in terms of total EM, EM-weighted $\log T$ and thermal width. If we choose the basis $\mathbf{B} = \mathbf{B}^{\rm Dirac} = \mathbf{I}$, the linear system (\ref{eqn:matrix_kb}) reduces to Eqs.~(\ref{eqn:linear_system}) and (\ref{eqn:linear_system_line2}).  For this choice of $\mathbf{B}$, the least L1-norm principle has a direct physical counterpart, namely a solution is sought such that the total EM is minimized. While the connection with a physical principle is appealing, this choice clearly gives an inferior inversion result \revagain{(see Figs.~\ref{fig:gaussian_test_dirac} and~\ref{fig:gaussian_curves_dirac}). The following point is also worth noting: Inversions performed using the basis given by Eq.~(\ref{eqn:basis_concat}) but with normalized Gaussians result in the same results as indicated in Figs.~\ref{fig:gaussian_test_dirac} and~\ref{fig:gaussian_curves_dirac} (i.e. the inversion scheme chooses to express the solution as a sum of isothermal components). This motivated us to use Gaussian basis functions that have maximum values of unity regardless of their widths. This effectively introduces a preference for broad solutions over narrow DEM solutions.} 

\revagain{There are likely better choices of basis functions for DEM inversions and the optimal set of basis functions (if it exists) may ultimately depend on the problem of interest. How to choose optimal choices of basis functions is an open question but theory and perhaps numerical simulations can provide guidance for future improvements. } 

\section{Dependence of DEM inversions results on the choice of the range for $\log T$}
\label{sec:tgrid}
\begin{figure}
\center\includegraphics[width=0.5\textwidth]{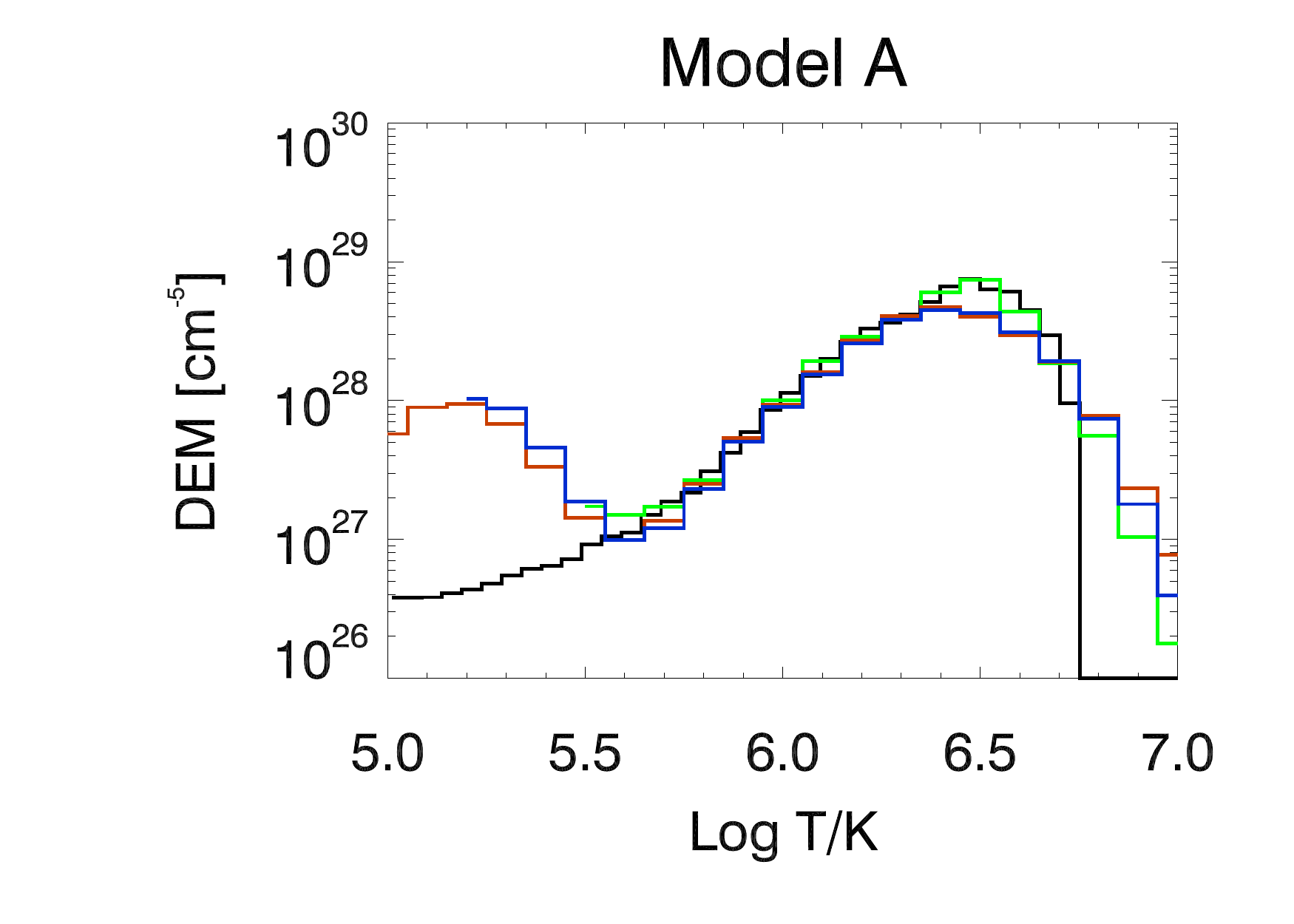}
\caption{\revision{Dependence of DEM solution with respect to choice of temperature grid. Plotted above are spatially-averaged DEM distributions of the model AR shown in Fig.~\ref{fig:anny_set1}. The black line indicates the underlying DEM model (model A, see section~\ref{sec:nlfff}). The green line indicates the sparse inversion solution for a $\log T/K$ grid spanning $5.5$ to $7.5$. The blue and red lines indicate solutions where the lower limit of the temperature grid is set to $\log T/K=5.2$ and $\log T/K = 5.0$, respectively. Of the three solutions, the one with $\log T/K = 5.5$ as the lower temperature limit (green line) shows the best match against the underlying model DEM over the temperature range $\log T/K \in [5.5, 6.7]$.}}\label{fig:tgrid}
\end{figure}

\revision{The DEM test results presented in this paper use a $\log T$ grid that spans $\log T/K = 5.5$ to $\log T/K = 7.5$ at intervals of $\Delta \log T/K = 0.1$. As Fig.~\ref{fig:response} indicates, however, some EUV channels of AIA (e.g. 335) also show significant response to plasma at $\log T/K < 5.5$. Our decision to restrict the DEM inversion to $\log T/K = 5.5$ is motivated by tests in which we varied the lower bound of the temperature grid. For example, Fig.~\ref{fig:tgrid} shows the spatially-averaged DEM distribution for model A of the test cases considered in section~\ref{sec:nlfff} (quasi-steady loop models by Malanushenko et al.). For generating synthetic AIA images (shown in the top row of Fig.~\ref{fig:anny_aia_images}), we used the model DEM values between $\log T/K  = 5.0$ to $\log T/K = 7.0$ (though there is no plasma in the model with $\log T/K > 6.8)$. We then performed pixel-by-pixel sparse DEM inversions on the synthetic AIA data and computed the spatial average of the DEM solutions.}

\revision{The black line in Fig.~\ref{fig:tgrid} shows the underlying model DEM. The green line shows the DEM solution for a $\log T/K$ grid spanning $5.5$ to $7.5$, with $\Delta \log T/K = 0.1$. The blue and red lines show the corresponding solutions when the lower bound of the temperature grid is $\log T /K = 5.2$ and $\log T/K = 5.0$, respectively. Between $\log T/K \approx 5.6$ and $\log T/K \approx 6.3$, all three solutions closely track the underlying model DEM. However, the blue and red solutions clearly show spurious enhancements of EM (relatively to the underlying model) at transition region temperatures ($\log T < 5.5$). Associated with this enhancement in the transition region is a deficit of EM between $\log T/K = 6.4$ and $\log T/K = 6.6$ (corresponding to the cores loops of the model AR). From this type of test, we decided to restrict the temperature grid to $\log T/K = 5.5$ and above for the validation exercises described in the paper. }

\end{document}